\documentclass[12pt]{article}
\usepackage[dvips]{graphicx}

\usepackage{graphics,epsfig}

\renewcommand{\bar}[1]{\overline{#1}}

\renewcommand{\bar}[1]{\overline{#1}}

\def\ru1{\rule[-0.4truecm]{0mm}{1truecm}}

\begin{document}

\begin{flushright}
CERN-PH-TH/2008-173
\end{flushright}


\centerline{\Large \bf Sivers Asymmetries for Inclusive Pion and
Kaon Production}
\centerline{\Large \bf in Deep-Inelastic Scattering}

\vspace{10mm}

\centerline{{\bf John Ellis$^{1}$, Dae Sung Hwang$^{2}$,}
and {\bf Aram Kotzinian$^{3,4,5}$}}

\vspace{5mm}

\vspace{4mm} \centerline{\it $^1$Theory Division, Physics Department,
CERN, 1211 Geneva 23, Switzerland}

\vspace{4mm} \centerline{\it $^2$Department of Physics, Sejong
University, Seoul 143--747, South Korea}

\vspace{4mm} \centerline{\it $^3$CEA DAPNIA/SPhN Saclay,
91191 Gif-sur-Yvette, France}

\vspace{4mm} \centerline{\it $^4$Joint Institute for Nuclear Reserch,
141980 Dubna, Moscow region, Russia}

\vspace{4mm} \centerline{\it $^5$Yerevan Physics Institute, 375036 Yerevan, Armenia}

\vspace{20mm}

\centerline{\bf Abstract}

\vspace{10mm}

\noindent
We calculate the Sivers distribution functions induced by
the final-state interaction due to one-gluon exchange in diquark
models of nucleon structure, treating the cases of scalar and axial-vector
diquarks with both dipole and Gaussian form factors.
We use these distribution functions to calculate the Sivers single-spin
asymmetries for inclusive pion and kaon production in deep-inelastic
scattering. We compare our calculations with the results of HERMES
and COMPASS, finding good agreement for $\pi^+$
production at HERMES, and qualitative agreement for $\pi^0$ and
$K^+$ production. Our predictions for pion and kaon production
at COMPASS could be probed with increased statistics.
The successful comparison of our calculations with the HERMES data
constitutes {\it prima facie} evidence that the quarks in the nucleon have some
orbital angular momentum in the infinite-momentum frame.

\vspace{0.5cm}

\noindent \vspace*{12mm}


\vspace{3mm}





\newpage


\section{Introduction}

\begin{figure}[h!]
\begin{center}
\includegraphics[width=0.75\columnwidth]{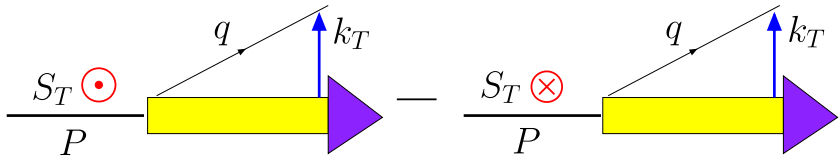}\\
\caption{\it Schematic depiction of the Sivers distribution function
$f_{1T}^{\perp}$. The spin vector $S_T$ of the nucleon points out of and
into the page, respectively, and $k_T$ is the transverse momentum of the extracted quark.}
\label{fig:siversdiagram}
\end{center}
\end{figure}

It is well-known that transverse-momentum-dependent distribution and
fragmentation functions, nowadays commonly referred to as TMDs, can have a
nontrivial spin dependences and that the so-called ``$T$-odd'' TMDs can lead to
single-spin asymmetries~\cite{Sivers,Collins:1992kk,Anselmino:1994tv,BM98}.
They are also often referred to as ``naively $T$-odd'',
because the appearance of these functions does not imply a violation of
time-reversal invariance, since they can arise through final-state
interactions. The Sivers distribution function $f_{1T}^\perp$,
schematically depicted in Fig.~\ref{fig:siversdiagram},
is the oldest example of such a function.

It describes the difference between the momentum distributions of quarks
inside protons transversely polarized in opposite directions.
The Sivers effect was put forward
as a possible explanation for the large
single-spin asymmetries observed in $p^\uparrow \, p \to \pi \, X$ experiments
\cite{Sivers,Anselmino:1994tv,Adams}.
Furthermore, it generates single-spin asymmetries in inclusive hadron
production in deep-inelastic scattering (SIDIS)~\cite{BM98,BHS}.
These have been measured by the HERMES collaboration to be nonzero
for $l\, p^\uparrow \to l'\, \pi \, X$~\cite{Airapetian:2004tw}, and
updated HERMES results for pion
and kaon production have been reported in~\cite{HERMESDIS2007}.
More recently, COMPASS has analyzed their SIDIS data on
pion and kaon production off a deutron target~\cite{COMPASS0802},
and collected data using a proton target in 2007.
Phenomenological analyses of SIDIS data have been performed in Refs.
\cite{Vogelsang05,Collins06,Arnold08,Anselmino0805}.
The Sivers effect may also result, e.g., in asymmetric di-jet correlations in
$p^\uparrow \, p \to \rm{jet} \, \rm{jet}\, X$~\cite{Boer:2003tx,Bacchetta:2005rm},
but these are not yet visible in the data analyzed to date~\cite{Abelev:2007ii}.

In recent years the importance in hadron physics of the
role of the transverse momenta of the partons
has been better recognized, since
they provide time-odd distribution and fragmentation functions,
and make possible single-spin asymmetries (SSA)
in hadronic processes~\cite{BM98,BHS,MT96}.
Specifically, it has been understood that one-gluon-exchange final-state interactions (FSI)
are a calculable mechanism for generating a transverse single-spin asymmetry
in SIDIS~\cite{BHS}.
This FSI generates a Sivers effect when the distribution functions are allowed to be functions of
the transverse momenta of the partons, as well as their longitudinal
momentum fractions.
Therefore, taking the transverse momenta of the partons into consideration
enlarges the realm of investigation of the nucleon structure.

A simple scalar diquark model was used in~\cite{BHS}
to demonstrate explicitly that this FSI can indeed give rise to
a leading-twist transverse SSA in SIDIS, which emerged from interference
between spin-dependent amplitudes for different
nucleon spin states.
It was observed in~\cite{BHS,weak} that the
same overlap integrals between light-cone wavefunctions
that describe the contribution to the nucleon anomalous magnetic moment from
a given quark flavor also appear in the Sivers distribution for
that quark flavor (with additional pieces in the integrand).
Since these integrals are the overlaps between light-cone wavefunctions
whose orbital angular momenta differ by $\Delta L^z = \pm 1$,
non-zero orbital angular momenta of the quarks inside the proton are
essential for the existence of the Sivers asymmetry~\cite{BHS,weak}.

In this paper we calculate the Sivers distribution functions in SIDIS induced by
the one-gluon exchange final-state interaction for models of the nucleon
in which the spectator diquark is treated as either a scalar or an axial-vector.
As we discuss below, the simplest SU(6) wavefunction of the nucleon suggests
that the spectator diquark would be in a combination of these two states.
We consider two types of form factor at the nucleon-quark-diquark vertices:
the dipole form factor used in~\cite{JMR97} and a Gaussian form factor.
When either of these form factors is used,
we find that at larger transverse momenta the asymmetry calculated using
one-gluon exchange may exceed unity in magnitude, indicating that unknown
higher-order effects must become important there~\footnote{A more complete description of
the FSI can be made by introducing an appropriate
Wilson-line phase factor in the definition of the distribution
functions of quarks in the nucleon~\cite{Sivers,collins,ji1,ji2,BMP03}.}. Imposing the
physical restriction that the asymmetry calculated by one-gluon exchange
not exceed unity changes rather little the Sivers single-spin
asymmetry after integration over the
transverse momentum, indicating that the results we present should be
quite reliable. The similarity of our predictions for dipole and Gaussian form factors
also indicates the stability of our results.

We compare our results for the Sivers single-spin asymmetries for
$\pi^+, \pi^0, \pi^-, K^+$ and $K^-$ with the SIDIS measurements made
by HERMES and COMPASS.
We find good agreement with the HERMES results for $\pi^+$
production, and qualitative agreement for $\pi^0$ and
$K^+$ production. The experimental errors in the HERMES measurements
of the Sivers asymmetries for $\pi^-$ and $K^-$ production do not permit
any definite conclusions to be drawn. The same is true of the current
measurements by COMPASS, and we look forward to increased statistics
that could further test our predictions for pion and kaon production in SIDIS.
However, the successful comparison of our calculations with the HERMES data
already constitutes {\it prima facie} evidence that the longitudinal projection
of the net quark angular momentum in the infinite-momentum frame is non-zero.

\section{Basic Formulae for the Sivers Asymmetry}

\subsection{Definition of the Sivers Asymmetry $A_{UT}^{\sin{(\phi_h-\phi_S)}}(x)$}

The SIDIS cross section $l+N^{\uparrow} \to l^\prime + h +x$ on
a transversely-polarized target contains 8 spin-dependent azimuthal modulations.
Here we consider only one of them -- the so-called Sivers asymmetry. The relevant
angular distribution of the cross section contains unpolarized (U) and Sivers (Siv) parts:
\begin{equation}
{d^6\sigma(x,y,z,P_T,\phi_h,\phi_S) \over dxdydzd^2P_Td\phi_S} = {d^6\sigma_U(x,y,z,P_T,\phi_h) \over dxdydzd^2P_Td\phi_S}+
S_T{d^6\sigma_{Siv}(x,y,z,P_T) \over dxdydzd^2P_Td\phi_S}\sin(\phi_h-\phi_S),
\end{equation}
where $\phi_h$ and $\phi_S$ are the azimuthal angles of the transverse momentum
of the produced hadron and the transverse spin of the target relative to the virtual photon direction.

The Sivers asymmetry is usually defined as

\begin{equation}
A_{UT}^{\sin(\phi_h-\phi_S)}(x,y,z,P_T)={2\over S_T}
{\int d\phi_h\int d\phi_S d^6\sigma(x,y,z,P_T,\phi_h,\phi_S)\sin(\phi_h-\phi_S)
\over \int d\phi_h\int d\phi_S d^6\sigma(x,y,z,P_T,\phi_h,\phi_S)},
\end{equation}
and we note that the integration singles out the $\sigma_U$ component
in the denominator and $\sigma_{Siv}$ in the numerator.

Within the LO QCD parton model, we have
\begin{equation}
{d^6\sigma_U(x,y,z,P_T,\phi_h) \over dxdydzd^2P_Td\phi_S}=C(x,y)
\Sigma_q e_q^2\ \int d^2{\bf k}_{\perp}^2\ f_{1}^q(x,{\bf k}_{\perp}^2)
D_{1q}^{h}(z,{\bf p}_\perp^2),
\label{sig_u}\\
\end{equation}
\begin{equation}
{d^6\sigma_{Siv}(x,y,z,P_T) \over dxdydzd^2P_Td\phi_S}=C(x,y)
\Sigma_q e_q^2\ \int d^2{\bf k}_{\perp}\
(-\, {|{\bf k}_{\perp}|\over M})\sin(\phi_q-\phi_S)\ f_{1T}^{\perp q}(x,{\bf k}_{\perp}^2) D_{1q}^{h}(z,{\bf p}_\perp^2),
\label{sig_siv}
\end{equation}
where
\begin{equation}\label{c}
    C(x,y)={\alpha^2_{em}\over 2ME}{1+(1-y)^2 \over xy^2},
\end{equation}
$f_1^q(x,{\bf k}_{\perp})$ and $f_{1T}^{\perp q}(x,{\bf k}_{\perp})$ are
the unpolarized and Sivers quark distribution
functions inside the nucleon, $\phi_q$ is the azimuthal angle of the active quark $q$,
$D_{1q}^{h}(z,{\bf p}_{\perp})$ is the unpolarized fragmentation function of $q$
into the hadron $h$, and
\begin{equation}\label{pperp}
    {\bf p}_\perp={\bf P}_T-z{\bf k}_{\perp}
\end{equation}
is the transverse momentum of the produced hadron with respect to
the fragmenting quark momentum.

In order to obtain the dependence of Sivers asymmetry on any single
kinematic variable such as $x$,
one must integrate the unpolarized and polarization-dependent parts of the
cross sections over the other three kinematic variables:
\begin{equation}
A_{UT}^{\sin(\phi_h-\phi_S)}(x)={\Delta {\hat \sigma}(x)
\over {\hat \sigma}(x)},
\label{siv}
\end{equation}
\begin{equation}
\Delta {\hat \sigma}(x)={\rm Int}\left [
C(x,y)\Sigma_q e_q^2\ \int d^2{\bf k}_{\perp}\
(-\, {{\bf k}_{\perp}\cdot{\bf P}_T\over M|{\bf P_T}|})\ f_{1T}^{\perp q}(x,{\bf k}_{\perp})\
D_{1q}^{h}(z,{\bf P}_T-z{\bf k}_{\perp})\right ] ,
\label{dsig}
\end{equation}
\begin{equation}
{\hat \sigma}(x)={\rm Int}\left [C(x,y)
\Sigma_q e_q^2\ \int d^2{\bf k}_{\perp}\ f_{1}^q(x,{\bf k}_{\perp})\
D_{1q}^{h}(z,{\bf P}_T-z{\bf k}_{\perp})\right ] \ ,
\label{sig}
\end{equation}
where ${\rm Int}\left [...\right ]$ denotes the following integration:
\begin{equation}
{\rm Int}\Big[\, G\, \Big]=\int_{z_{\rm min}}^{z_{\rm max}}dz \int_{P_{T\, {\rm min}}}^{P_{T\, {\rm  max}}} d|{\bf P_T}||{\bf P_T}| \int_0^{2\pi} d \phi
\int_0^{k_{\perp\, {\rm max}}} d|{\bf k}_{\perp}||{\bf k}_{\perp}|\Big[\, G\, \Big].
\label{int}
\end{equation}
The derivation of (\ref{dsig}) is given in the Subsection \ref{s:kt-int}.

In the later comparisons with experimental data,
we use the following integration limits for asymmetries on a proton target at HERMES:
$0.2<z<0.8$, $0.05$ GeV $< |{\bf P_T}| < 1.2$ GeV, and for asymmetries
on a deuteron target at COMPASS we use $0.2<z<0.8$, $0.1$ GeV $< |{\bf P_T}| < 2.0$ GeV,
corresponding to the kinematic conditions of these experiments.

\subsection{Intrinsic ${\bf k}_{\perp}$ Integration}\label{s:kt-int}

Let us consider two integrals:
\begin{equation}
R_1=\int d^2{\bf k}_{\perp}\
f({\bf k}_{\perp}^2) D({\bf p}_\perp^2),
\label{int1}
\end{equation}
\begin{equation}
R_2=\int d^2{\bf k}_{\perp}\
{|{\bf k}_{\perp}|\over M}\sin(\phi_q-\phi_S)\ f({\bf k}_{\perp}^2) D({\bf p}_\perp^2).
\label{int2}
\end{equation}
The integrand of $R_1$ is a scalar function of vectors ${\bf k}_{\perp}$ and ${\bf P}_T$ in
the two-dimensional transverse momentum space. This means that $R_1$ can be a function
only of $z$ and ${\bf P}_T^2$:
$$R_1=r_1(z,{\bf P}_T^2).$$
On the other hand, (\ref{int2}) can be represented as
\begin{equation}
R_2={\hat S}_1F_2-{\hat S}_2F_1 \ ,
\label{int2-1}
\end{equation}
Where ${\hat S}_{1,2}$ are the components of the two-dimensional transverse spin vector
${\bf{\hat S}}$ and the $F_i$ are the components of
\begin{equation}
{\bf F}=\int d^2{\bf k}_{\perp}\
{{\bf k}_{\perp}\over M}\ f({\bf k}_{\perp}^2) D({\bf p}_\perp^2).
\label{int2-2}
\end{equation}
Since ${\bf F}$ is a two-dimensional vector, and the only vector remaining after integration
in (\ref{int2-2}) is ${\bf P}_T$, we have
\begin{equation}
{\bf F}={\bf P}_{T}\Phi({\bf P}_{T}^2),
\label{int2-3}
\end{equation}
and hence
\begin{equation}
\Phi({\bf P}_{T}^2)={{\bf P}_T \cdot {\bf F} \over {\bf P}_{T}^2}.
\label{int2-4}
\end{equation}
From Eqs. (\ref{int2-1} -- \ref{int2-4}) we finally obtain
$$R_2=\sin(\phi_h-\phi_S)r_2(z,{\bf P}_T^2),$$
where
\begin{equation}
r_2(z,{\bf P}_T^2)=\int d^2{\bf k}_{\perp}\
{{\bf k}_{\perp}\cdot{\bf P}_T\over M|{\bf P_T}|}\ f({\bf k}_{\perp}^2) D({\bf p}_\perp^2).
\label{int2f}
\end{equation}

\subsection{Calculations of the Sivers Function in Diquark Models}\label{sec:siversfunction}

\begin{figure}[h!]
\begin{center}
\includegraphics[width=0.75\columnwidth]{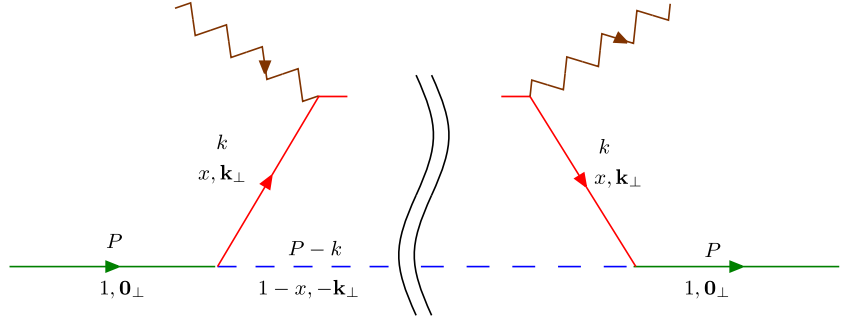}\\
\caption{\it Leading-order diagram contributing to $f_1$ in a diquark model.}
\label{fig:f1diagram}
\end{center}
\end{figure}

\begin{figure}[h!]
\begin{center}
\includegraphics[width=0.75\columnwidth]{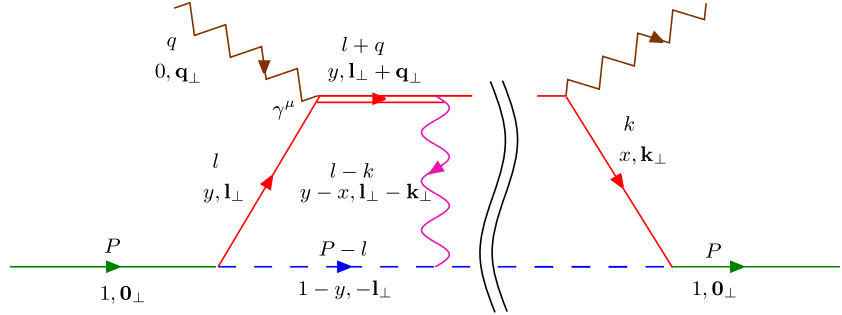}\\
\caption{\it Leading-order diagram contributing to the Sivers function $f_{1T}^{\perp}$ in
a diquark model, via a one-gluon exchange final-state interaction (FSI).}
\label{fig:f1ptdiagram}
\end{center}
\end{figure}

In this subsection we repeat the derivation of the Sivers function in the
scalar diquark model with a constant Yukawa vertex
given in~\cite{BHS, BBH03},
and later generalize the result to the scalar and axial-vector diquark models with
non-trivial form factors $g(k^2)$ at the nucleon-quark-diquark vertex.
We obtain, from (\ref{sn2}) in the Appendix of this paper or from (22) of~\cite{BBH03},
the following expression for the distribution function in the Yukawa model:
\begin{equation}
f_1(x, {\bf k}_{\perp})=
g^2(1-x){{\bf k}_{\perp}^2+(xM+m)^2\over ({\bf k}_{\perp}^2+B)^2}
\ .
\label{app0}
\end{equation}
From Eqs. (28) and (30) of \cite{BBH03}, we also have
\begin{eqnarray}
k_{\perp}^x\, f_{1T}^{\perp}(x, {\bf k}_{\perp})&=&
-g^2{e_1e_2\over 4\pi}\
(xM+m)(1-x)\, {1\over ({\bf k}_{\perp}^2+B)}
\nonumber\\
&&\times\ {1\over \pi}\,
\int d^2{\bf l}_{\perp}\,
{1\over ({\bf l}_{\perp}^2+B)}\
{(l_{\perp}-k_{\perp})^x \over
[({\bf l}_{\perp}-{\bf r}_{\perp})^2+{\lambda}_g^2]}\ .
\label{app1}
\end{eqnarray}
Here we set to unity the parameter $a$ in Eq. (30) of~\cite{BBH03}.
Using (\ref{app0}) and (\ref{app1}), we can write
\begin{equation}
{f_{1T}^{\perp}(x,{\bf k}_{\perp})\over f_1(x,{\bf k}_{\perp})}=
{e_1e_2\over 4\pi}\, {M(xM+m)\over (xM+m)^2+{\bf k}_{\perp}^2}\,
k_{\perp}^x\, R(x,{\bf k}_{\perp})\ ,
\label{Sfration1aa}
\end{equation}
where $R(x,{\bf k}_{\perp})$ is given by
\begin{equation}
k_{\perp}^x\, R(x,{\bf k}_{\perp})\times {1\over {\bf k}_{\perp}^2+B}
={-1\over \pi}\,
\int d^2{\bf l}_{\perp}\,
{1\over ({\bf l}_{\perp}^2+B)}\
{(l_{\perp}-k_{\perp})^x \over
[({\bf l}_{\perp}-{\bf r}_{\perp})^2+{\lambda}_g^2]} \; \equiv \; I\ .
\label{R1aa}
\end{equation}
We calculate the RHS of (\ref{R1aa}) by first taking the direction of ${\bf k}_{\perp}$ as
the $x$ direction and using $k$ and $l$ to denote
$|{\bf k}_{\perp}|$ and $|{\bf l}_{\perp}|$, respectively, so that
\begin{eqnarray}
{\bf k}_{\perp}&=&k(1,0) ,
\nonumber\\
{\bf l}_{\perp}&=&l({\rm cos}\phi ,{\rm sin}\phi ) ,
\nonumber\\
(l_{\perp}-k_{\perp})^x&=&l{\rm cos}\phi -k ,
\nonumber\\
({\bf l}_{\perp}-{\bf k}_{\perp})&=&(l{\rm cos}\phi -k, l{\rm sin}\phi ) ,
\nonumber\\
({\bf l}_{\perp}-{\bf k}_{\perp})^2&=&l^2+k^2-2kl{\rm cos}\phi \ .
\label{app3}
\end{eqnarray}
Then, $I$ in (\ref{R1aa}) becomes
\begin{eqnarray}
I&=&{-1\over \pi}\,
\int ldld\phi\ {l{\rm cos}\phi -k\over
(l^2+B)(l^2+k^2-2kl{\rm cos}\phi )}
\nonumber\\
&=&
\int_0^{\infty} ldl\ {1\over (l^2+B)(l^2+k^2)}\
\Big( {-1\over \pi}\, \int_0^{2\pi} d\phi\
{l{\rm cos}\phi -k\over 1-{2kl\over l^2+k^2}{\rm cos}\phi }\Big)
\ .
\label{app4}
\end{eqnarray}
After some calculations, we find that the last factor in (\ref{app4})
gives
\begin{equation}
\Big( {-1\over \pi}\, \int_0^{2\pi} d\phi\
{l{\rm cos}\phi -k\over 1-{2kl\over l^2+k^2}{\rm cos}\phi }\Big)
=
\left\{ \begin{array}{ll}
2\, {l^2+k^2\over k}&\mbox{when $l < k$}\\
0               &\mbox{when $l > k$}
\end{array}\right. \label{t2aa}
\end{equation}
Using (\ref{t2aa}), the expression (\ref{app4}) becomes
\begin{eqnarray}
I&=&\int_0^k ldl\ {1\over (l^2+B)(l^2+k^2)}\
2{l^2+k^2\over k}
\nonumber\\
&=&
{1\over k}\ \int_0^k 2ldl\ {1\over (l^2+B)}
\nonumber\\
&=&
{k_{\perp}^x\over {\bf k}_{\perp}^2}\ \int_0^{{\bf k}_{\perp}^2}
d({\bf l}_{\perp}^2)\ {1\over ({\bf l}_{\perp}^2+B)}\ ,
\label{app5}
\end{eqnarray}
where we use ${1 / k}={k_{\perp}^x / {\bf k}_{\perp}^2}$, since
we take the direction of ${\bf k}_{\perp}$ as the $x$ direction.
From (\ref{R1aa}) and (\ref{app5}), we have
\begin{equation}
R(x,{\bf k}_{\perp})\times {1\over {\bf k}_{\perp}^2+B}
={1\over {\bf k}_{\perp}^2}\ \int_0^{{\bf k}_{\perp}^2}
d({\bf l}_{\perp}^2)\ {1\over ({\bf l}_{\perp}^2+B)}\ .
\label{app6}
\end{equation}
Then, using $R(x,{\bf k}_{\perp})$ in (\ref{app6}), we can obtain the
Sivers function from the formula (\ref{Sfration1aa}).
The result is identical to the result in~\cite{BHS,BBH03}.

Eq. (\ref{app6}) is for the Yukawa model, presented in the Appendix, in which
the nucleon-quark-diquark vertex
is $g(k^2)=1$.
We can generalize (\ref{app6}) for a general form factor
$g(k^2)=g(x,{\bf k}_{\perp})$ by using the following formula:
\begin{equation}
R(x,{\bf k}_{\perp})\times {g(x,{\bf k}_{\perp})\over {\bf k}_{\perp}^2+B}
={1\over {\bf k}_{\perp}^2}\ \int_0^{{\bf k}_{\perp}^2}
d({\bf l}_{\perp}^2)\ {g(x,{\bf l}_{\perp})\over ({\bf l}_{\perp}^2+B)}\ .
\label{app6gen}
\end{equation}
Using formula (\ref{app6gen}), we now calculate $R(x,{\bf k}_{\perp})$ for
the dipole and Gaussian form factors at the proton-quark-diquark vertex,
respectively, obtaining the results shown in
(\ref{R1dp}) and (\ref{R1gaussian}).

\subsection{Generalized Diquark Models}

Whilst the simplest possibility for the spectator diquark is the scalar case,
even in the absence of orbital angular momentum in the nucleon rest frame,
it may exist in a spin-one, axial-vector state. Indeed, this possibility is
realized in the simplest non-relativistic SU(6) wavefunction for the proton:
\begin{eqnarray}
|p\uparrow \rangle&=&{1\over {\sqrt{3}}}\Big( {\sqrt{2\over 3}}|uu\ 1+1\rangle|d\downarrow \rangle
-{\sqrt{1\over 3}}|uu\ 10\rangle|d\uparrow \rangle\Big)
\label{su6p1}\\
&&-{1\over {\sqrt{6}}}\Big( {\sqrt{2\over 3}}|ud\ 1+1\rangle|u\downarrow \rangle
-{\sqrt{1\over 3}}|ud\ 10|u\uparrow \rangle\Big)
+{1\over {\sqrt{2}}}|ud\ 00>|u\uparrow \rangle
\nonumber\\
|p\downarrow \rangle&=&{1\over {\sqrt{3}}}\Big( -{\sqrt{2\over 3}}|uu\ 1-1\rangle|d\uparrow \rangle
+{\sqrt{1\over 3}}|uu\ 10\rangle|d\downarrow \rangle\Big)
\label{su6p2}\\
&&-{1\over {\sqrt{6}}}\Big( -{\sqrt{2\over 3}}|ud\ 1-1>|u\uparrow \rangle
+{\sqrt{1\over 3}}|ud\ 10\rangle|u\downarrow \rangle\Big)
+{1\over {\sqrt{2}}}|ud\ 00\rangle|u\downarrow \rangle
\ .
\nonumber
\end{eqnarray}
In each of the terms in (\ref{su6p1}) and (\ref{su6p2}), we have exhibited
the spin states of the spectator diquark, showing explicitly that it is in a combination
of scalar and axial-vector configurations in different spin states.

If the proton does have a naive SU(6)
wavefunction in its rest frame, the
distribution functions of the $u$ and $d$ quarks inside the proton, $f_1^u$ and $f_1^d$,
are given by
\begin{equation}
f_1^u={3\over 2}f_1^s+{1\over 2}f_1^a\ ,\qquad f_1^d=f_1^a\ ,
\label{pr}
\end{equation}
and those inside the deuteron are given by
\begin{equation}
f_{{\rm deu}\, 1}^u=f_{{\rm deu}\, 1}^d=f_1^u+f_1^d\ .
\label{de}
\end{equation}
The same relations also hold for the Sivers distribution functions,
as we discuss below in more detail.

In the following, we allow for non-trivial
form factors at the nucleon-quark-diquark vertices,
in both the scalar ($s$) and axial-vector ($a$) diquark cases:
\begin{equation}
\Upsilon_s=g_s(k^2)\ , \qquad
\Upsilon_a^\mu ={g_a(k^2)\over {\sqrt{3}}}\gamma^\mu\gamma_5 \ ,
\label{sa1aa}
\end{equation}
Two specific models for the form factors $g_s(k^2)$ and $g_a(k^2)$
are discussed in the following Sections.

\section{Dipole Form Factor}\label{sec:dipole}

\subsection{Calculations}

We first consider a dipole model for the nucleon-quark-diquark vertex:
\begin{equation}
g_{R}(k^2)={\sqrt{N_{R}}}\ \, {k^2-m^2\over (k^2-\Lambda^2)^2}
={\sqrt{N_{R}}}\ \, {(k^2-m^2)(1-x)^2\over ({\bf k}_{\perp}^2+B_{R})^2}\ ,
\label{sa2aa}
\end{equation}
where $R=s$ for the scalar and $R=a$ for the axial-vector diquark,
we used~\cite{JMR97,BBH03}
\begin{equation}
-k^2(x, {\bf k}_{\perp}^2)={{\bf k}_{\perp}^2\over 1-x}+{x\over 1-x}M_R^2-xM^2 ,
\label{g2}
\end{equation}
and $B_{R}$ is given by
\begin{equation}
B_{R}=(1-x)\Lambda^2+xM_{R}^2-x(1-x)M^2\ .
\label{sa3aa}
\end{equation}
We then have the following distribution functions:
\begin{equation}
f_1^R(x,{\bf k}_{\perp})=
{N_R \over 16 \pi^3}{(1-x)^3[(xM+m)^2+{\bf k}_{\perp}^2]\over ({\bf k}_{\perp}^2+B_{R})^4}\ ,
\label{sa4}
\end{equation}
yielding the following when integrated over ${\bf k}_{\perp}$:
\begin{equation}
f_1^R(x)=
{N_R \over 16 \pi^3}{\pi(1-x)^3[2(xM+m)^2+B_{R}]\over 6B_{R}^3}\ .
\label{sa4int}
\end{equation}
The normalization factors $N_R$ are fixed from the condition
\begin{equation}
\int_0^1 dx f_1^R(x)= 1\ .
\label{normsa}
\end{equation}
In the present cases of dipole form factors, we obtain from (\ref{app6gen})
\begin{eqnarray}
R(x,{\bf k}_{\perp})\times {1\over ({\bf k}_{\perp}^2+B_{\Lambda})^2}
&=&{1\over {\bf k}_{\perp}^2}\,
\int_0^{{\bf k}_{\perp}^2}d({\bf l}_{\perp}^2)\,
{1\over ({\bf l}_{\perp}^2+B_{\Lambda})^2}
\nonumber\\
&=&{1\over {\bf k}_{\perp}^2}\,
(-{1\over {\bf k}_{\perp}^2+B_{\Lambda}}+{1\over B_{\Lambda}})
={1\over B_{\Lambda}({\bf k}_{\perp}^2+B_{\Lambda})}\ .
\label{R1dp}
\end{eqnarray}
Thus, we obtain finally the Sivers distribution functions
for the dipole form factors as follows:
\begin{equation}
f_{1T}^{\perp R}(x,{\bf k}_{\perp})=a_R\,
{e_1e_2\over 4\pi}\
{N_R(1-x)^3M(xM+m)\over 16{\pi}^3B_R({\bf k}_{\perp}^2+B_{R})^3}\ ,
\label{sa6}
\end{equation}
In deriving (\ref{sa6}), we use for the gauge-field coupling to the axial-vector diquark
in Fig. \ref{fig:f1ptdiagram} the simple form
$i e_2\, g^{\alpha\beta}\, ((P-l)+(P-k))^\mu$, which is equivalent,
for each polarization state, to the gauge-field coupling to a scalar
diquark~\footnote{
We note that the results for the axial-vector diquark in (\ref{sa4}),
(\ref{sa4int}) and (\ref{sa6}) are different from those of~\cite{BSY04},
which are
obtained if $\sum_{\lambda} {\epsilon}_{\mu}^{(\lambda)*} {\epsilon}_{\nu}^{(\lambda)}
= - g_{\mu\nu}$ is used.
Ref. \cite{BCR0807} considers various different possibilities for the polarization sum
of the axial-vector diquark.
More general forms of gauge-field coupling to the axial-vector diquark were used
in Refs. \cite{BSY04,BCR0807,GGS07}.}.
Following~\cite{BHS}, we fix ${e_1e_2\over 4\pi}=-C_F\alpha_S$, where
$C_F = {4\over 3}$.

The values of $a_R$ are given by the overlaps of the proton wave functions
of positive and negative helicities.We find:
\begin{equation}
\label{avalues}
a_s=1 \; \; \; {\rm and} \; \; \;  a_a=-{1\over 3}.
\end{equation}
Details of the derivation are presented in the Appendix, where the relations between
the SU(6) and light-cone wavefunctions are discussed, as well as the relations between
light-cone and Bjorken-Drell spinors.



We use $\alpha_S\approx 0.3$ and
choose the following values for the parameters of the model
studied in this Section:
\begin{equation}
m=0.36\ {\rm GeV},\ \ \ M_s=0.6\ {\rm GeV},\ \ \ M_a=0.8\ {\rm GeV},
\ \ \ \Lambda =0.65\ {\rm GeV}.
\label{sa8}
\end{equation}
We assume a Gaussian transverse-momentum dependence for the fragmentation
functions:
\begin{equation}
D_{1q}^{h}(z,{\bf p}_\perp^2)=
{1\over \pi \mu_2^2} \ e^{-{{\bf p}_\perp^2}\over \mu_2^2}\,
D_{1q}^h(z)\ ,
\label{FFzp}
\end{equation}
with $\mu_2^2=0.2$ GeV$^2$ as obtained in \cite{Anselmino:2005nn}.
We use the leading-order fragmentation functions of~\cite{dss} for the integrated fragmentation functions $D_{1q}^h(z)$ . The ${\bf k}_{\perp}$ integration is performed as described in Section~\ref{s:kt-int}.

\begin{figure}
\begin{center}
\includegraphics[width=1.\columnwidth]{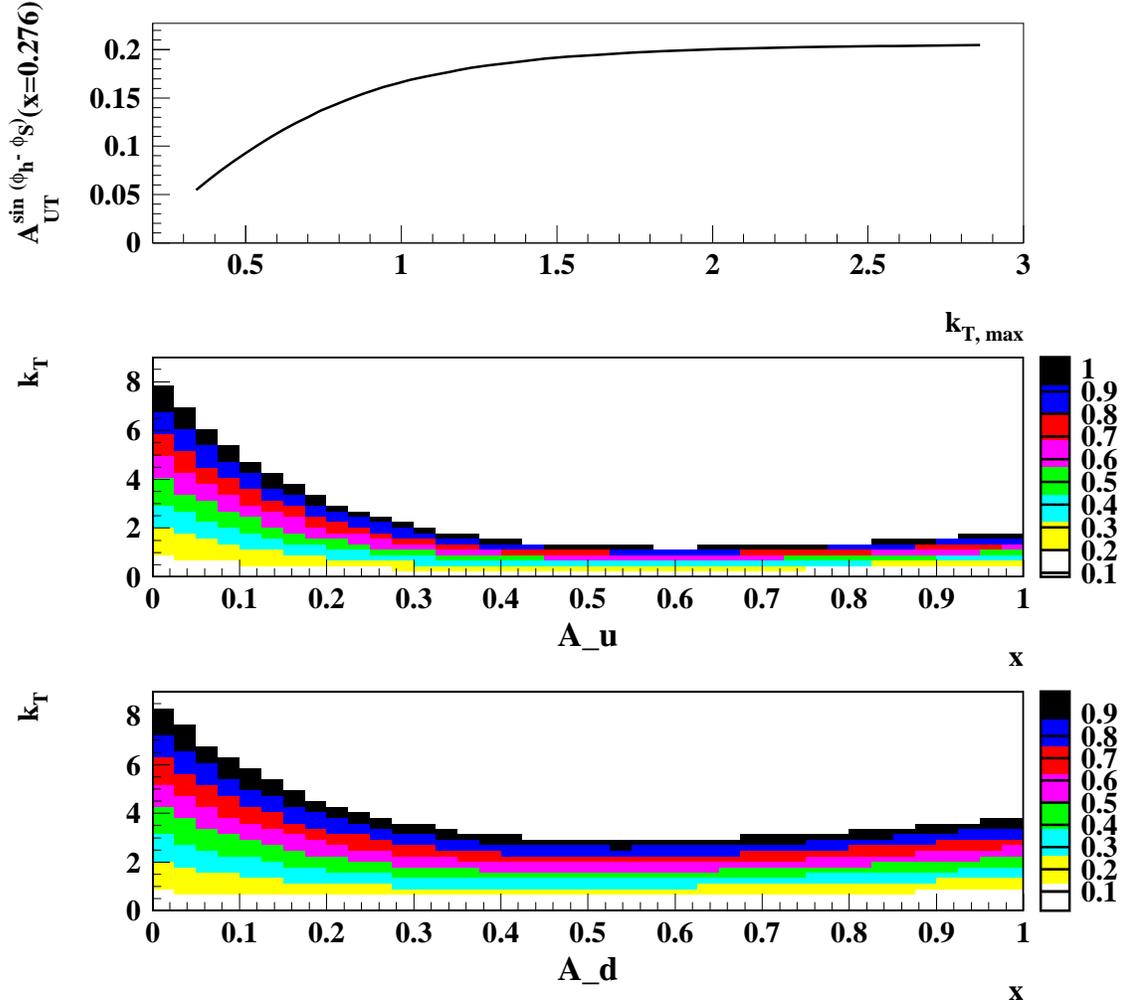}
\end{center}
\caption{\it In the top panel,
we show the Sivers asymmetry for $\pi^+$ production off proton target
for HERMES kinematics at $x=0.276$ as a function of the upper
limit of the $|{\bf k}_{\perp}|$ integration.
The analyzing power of the Sivers function for scattering off a $u$ quark (middle panel) and
a $d$ quark (bottom panel) as functions of $x$ and $k_T=|{\bf k}_{\perp}|$. The calculated
values exceed the unitarity limit in the white regions at larger $x$ and $k_T$.
All these plots are obtained using dipole form factors.}
\label{Fig:kmax_dip}
\end{figure}

We present in the top panel of Fig.~\ref{Fig:kmax_dip} the dependence of the calculated asymmetry
on the upper limit in the ${\bf k}_{\perp}$ integral.
As one can see, the saturation of integral takes place around 2 GeV,
which is a rather high value, though about 90~\% of the integral is provided by the region
$|{\bf k}_{\perp}| < 1$~GeV. For definiteness, in this Section using dipole form factors,
we use $|{\bf k}_{\perp}|_{\rm max}=2.5$ GeV for the upper limit of the ${\bf k}_{\perp}$ integration.
The leading-order perturbative QCD approach is believed to be applicable to SIDIS
when all transverse momenta are much smaller than the virtuality of the hard
scattering, $Q$. However, in SIDIS at fixed energy there is a strong correlation between the
mean values of $x$ and $Q^2$: for example, at HERMES $\langle Q^2\rangle(x=0.18)
\approx 4$  GeV and $\langle Q^2\rangle(x=0.28) \approx 6$ GeV.
Thus the highest virtuality at HERMES corresponding to the last populated $x$ bins is
of the same order
of magnitude as the saturation value for the intrinsic transverse momentum integration.
This consideration, together with the fact that the simple
diquark model used here treats only the valence quarks,
shows that this approach cannot be considered reliable for low values of $x$. Hence, our
results should be considered as applicable only to $x > 0.1$.

Another issue bearing on the accuracy of our results is that unitarity requires
the analyzing power of the Sivers function to be less than unity in modulus
for all values of $x$ and ${\bf k}_{\perp}$:
$A_{u,d} \equiv |k_{\perp}f_{1T}^{\perp u,d}(x,{\bf k}_{\perp})/Mf_{1}^{u,d}(x,{\bf k}_{\perp})| \leq 1$.
This is not always the case for the simplified leading-order one-gluon-exchange FSI
that we consider, which tends to yield larger values at large $x$ and
${\bf k}_{\perp}$, as seen in the middle panel of
Fig.~\ref{Fig:kmax_dip} for scattering off a $u$ quark, and in the bottom panel for
scattering off a $d$
quark~\footnote{A related issue is that the asymmetry we calculate is
large close to the contour in the $(x, {\bf k}_{\perp})$ beyond which unitarity is violated.}.
The issue would be resolved if a full higher-order calculation were performed, but this
is currently not available, so one must estimate the error incurred by including
unphysical values in the integration over ${\bf k}_{\perp}$.
In fact, we find that the naive predictions
obtained by ignoring the unitarity issue differ from the modified calculations,
that are obtained by truncating the
${\bf k}_{\perp}$ integration at the $x$-dependent unitarity limit shown in
Fig.~\ref{Fig:kmax_dip}, by far less
than the present experimental uncertainties, so this theoretical error may be neglected
for the time being.

\begin{figure}
\includegraphics[width=1.\columnwidth]{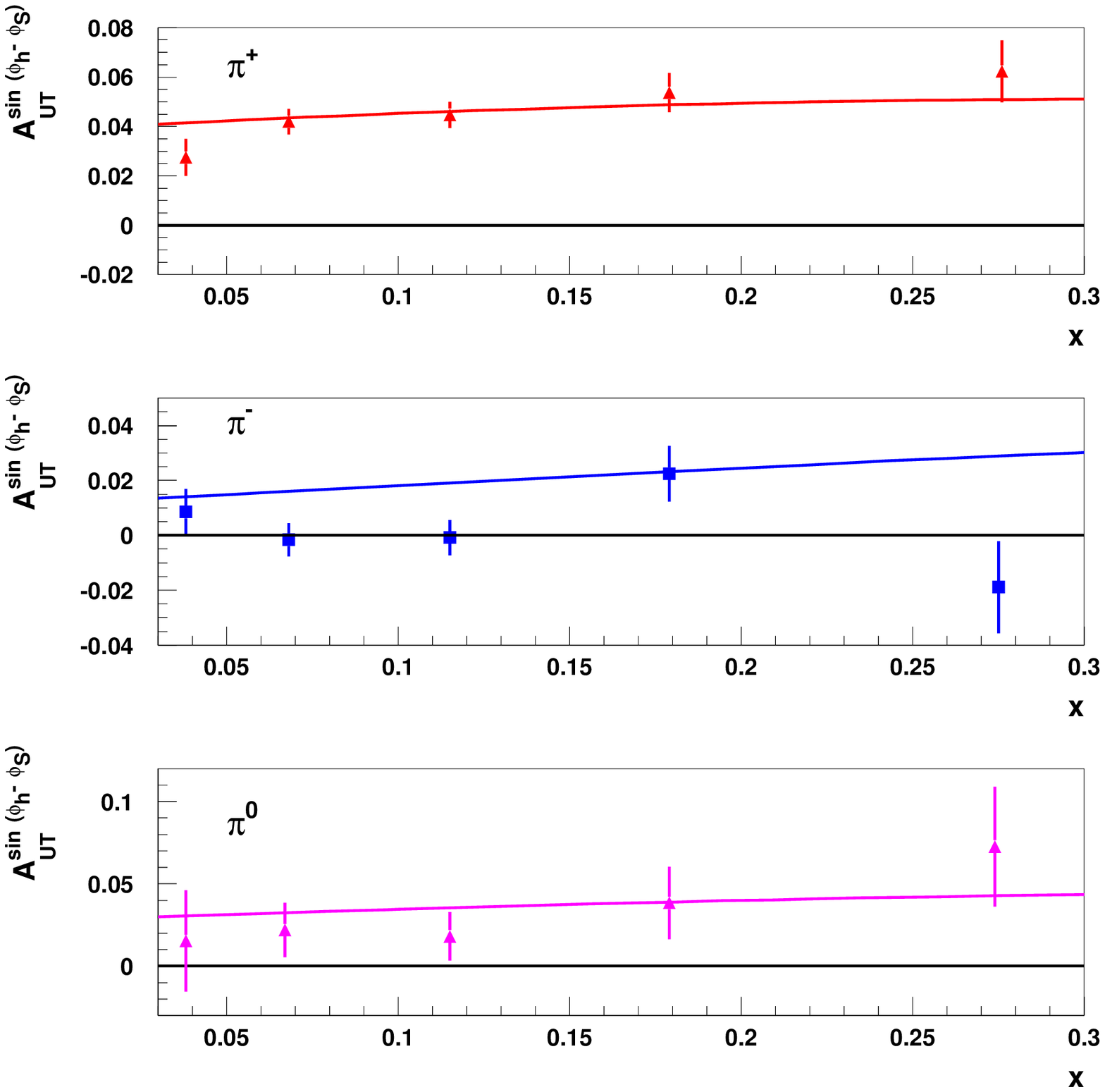}
\caption{\it Comparisons of our predictions for the Sivers asymmetries in the
production of $\pi^+$ (top panel), $\pi^-$ (middle panel) and $\pi^0$
(bottom panel) with HERMES data, assuming $\alpha_s=0.3$ and $\Lambda=0.65$ GeV
for the dipole model parameters.}
\label{Fig:h_dip0306}
\end{figure}

\begin{figure}
\includegraphics[width=1.\columnwidth]{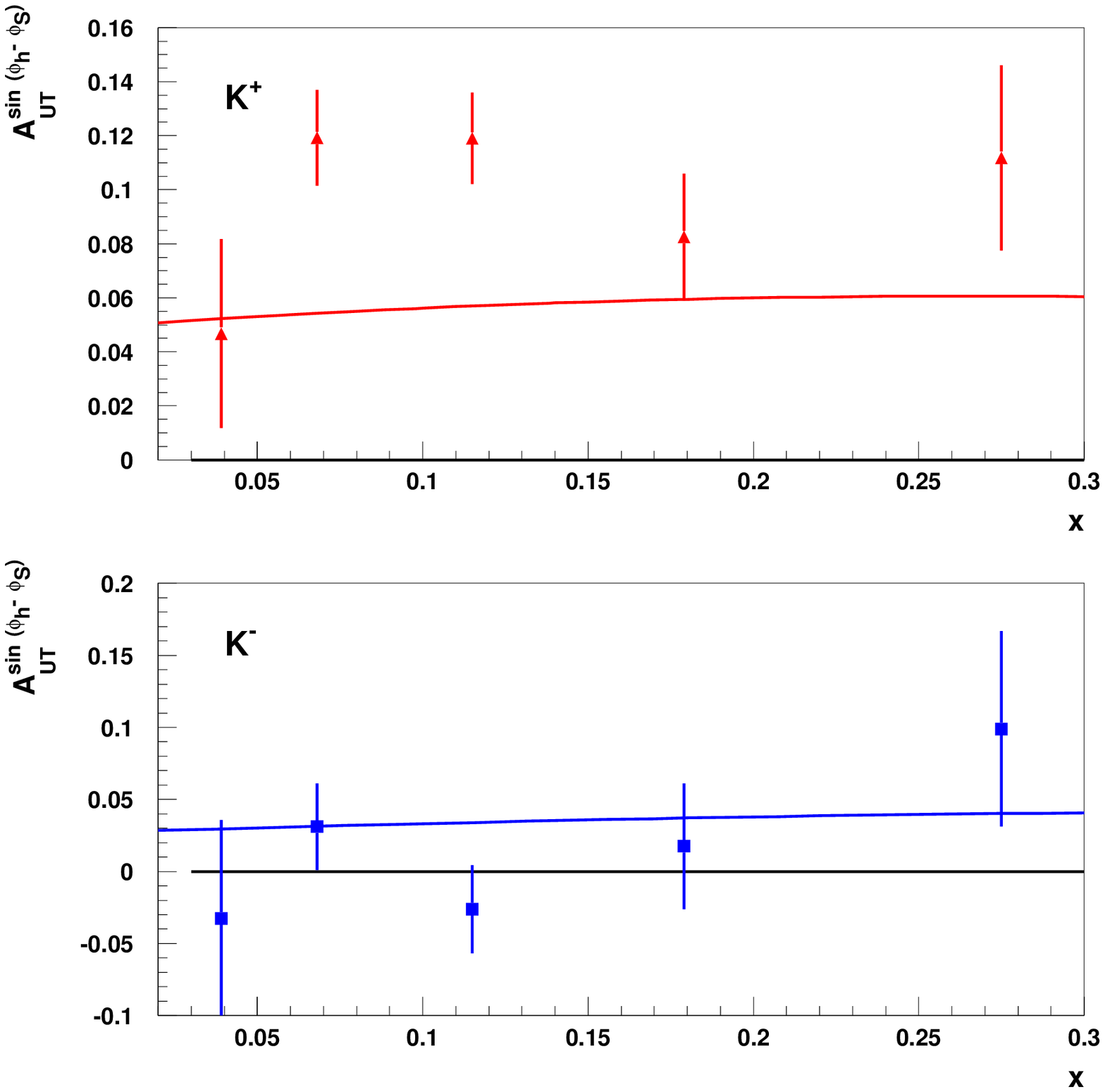}
\caption{\it Comparisons of our predictions for the Sivers asymmetries in the
production of $K^+$ (top panel) and $K^-$
(bottom panel) with HERMES data, assuming $\alpha_s=0.3$ and $\Lambda=0.65$ GeV
for the dipole model parameters.}
\label{Fig:h_kdip0306}
\end{figure}

\begin{figure}
\includegraphics[width=1.\columnwidth]{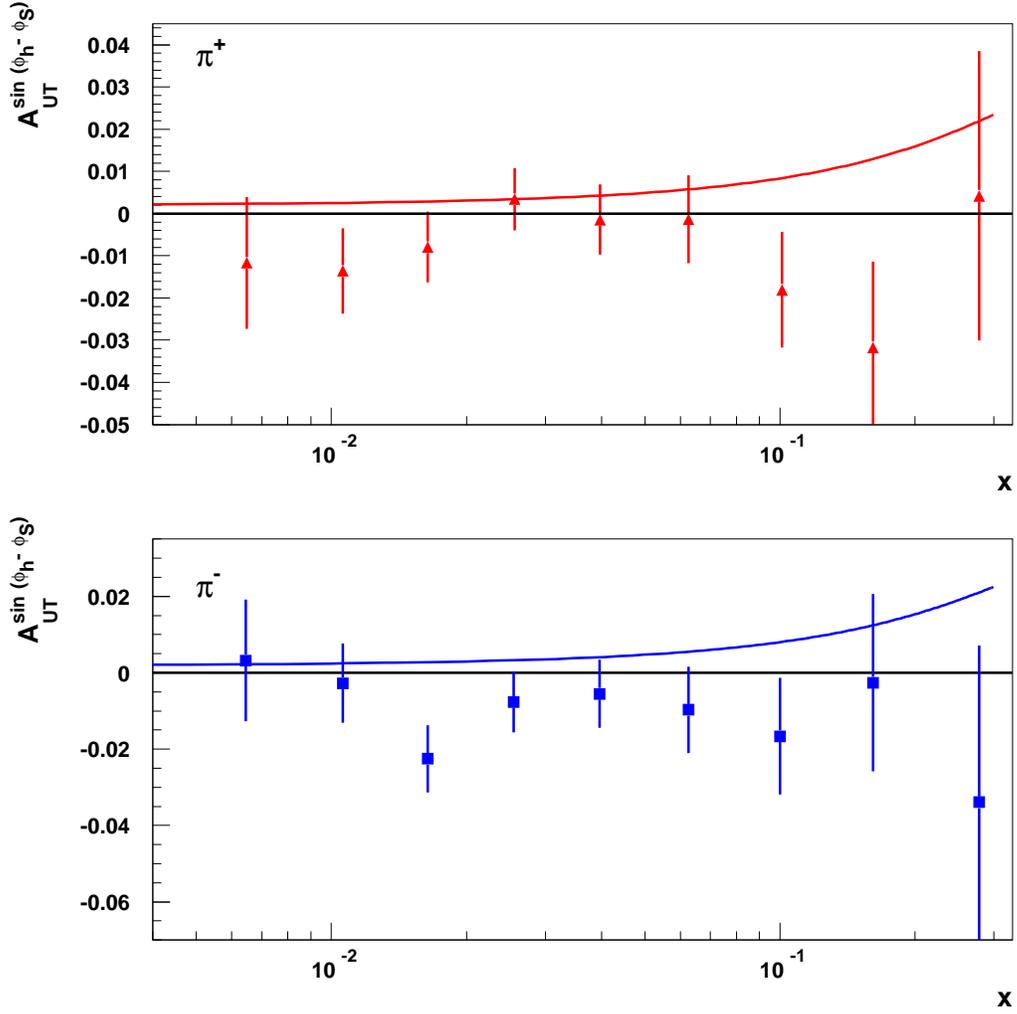}
\caption{\it Comparisons of our predictions for the Sivers asymmetries in the
production of $\pi^+$ (top panel) and $\pi^-$ (bottom panel) with COMPASS deuteron target data,
assuming $\alpha_s=0.3$ and $\Lambda=0.65$ GeV
for the dipole model parameters.}
\label{Fig:c_dip0306}
\end{figure}

\begin{figure}
\includegraphics[width=1.\columnwidth]{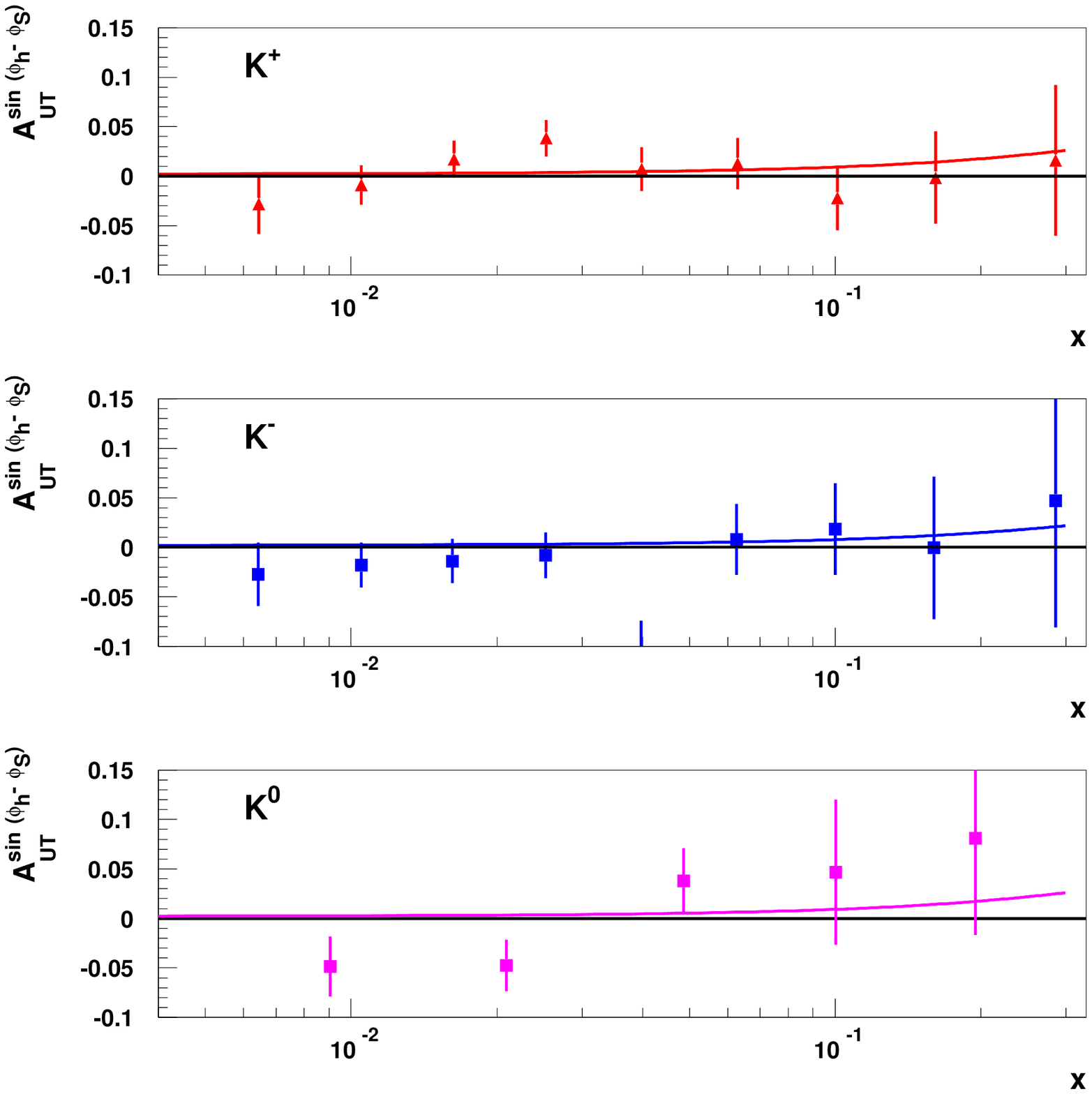}
\caption{\it Comparisons of our predictions for the Sivers asymmetries in the
production of $K^+$ (top panel), $K^-$ (middle panel) and $K^0$
(bottom panel) with COMPASS deuteron target data, assuming $\alpha_s=0.3$ and $\Lambda=0.65$ GeV for the dipole model parameters.}
\label{Fig:c_kdip0306}
\end{figure}

\subsection{Comparisons with Data}

In Figs. \ref{Fig:h_dip0306} and \ref{Fig:h_kdip0306} we compare
the Sivers asymmetries for pion and kaon production calculated with
a dipole form factor using $\alpha_s=0.3$ and $\Lambda=0.65$ GeV
with data from the HERMES Collaboration obtained using a proton target.
We see that our predictions for the $\pi^+$ asymmetry (top panel of
Fig.~\ref{Fig:h_dip0306}) agree very well
with the HERMES data, which exhibit a relatively significant positive asymmetry.
The HERMES data for the $\pi^-$ asymmetry (middle panel) are more equivocal, though they
are compatible with positive values that are smaller than for $\pi^+$, as
predicted by our calculations. We also predict a positive asymmetry for $\pi^0$
production, which is in qualitative agreement with the HERMES data, as seen in
the bottom panel of Fig.~\ref{Fig:h_dip0306}. In the case of the HERMES data
for kaons, we see qualitative agreement for the $K^+$ asymmetry shown in the
top panel of Fig.~\ref{Fig:h_kdip0306}, though the experimental values are somewhat
larger than the predictions, albeit with large errors.
This difference could be explained by the fact that we do not include the ${\bar s}$
contribution, which dominates in fragmentation to the $K^+$. In the case of the $K^-$
asymmetry shown in the lower panel of Fig.~\ref{Fig:h_kdip0306}, the experimental
values are similar to the predictions, though again with relatively large errors.

The corresponding comparisons for COMPASS data on pion and kaon
production are shown in Figs.~\ref{Fig:c_dip0306} and \ref{Fig:c_kdip0306},
respectively. In this case, again using a dipole form factor and $\alpha_s=0.3$ and
$\Lambda=0.65$~GeV, we predict small asymmetries for all three
charge states for both pions and kaons. These predictions are
compatible with the data available from COMPASS for $\pi^\pm, K^\pm$ and $K^0$
production.

We conclude that our model gives qualitatively successful predictions for
the Sivers asymmetries if a dipole form factor is assumed.


\section{Gaussian Form Factor}\label{sec:gaussian}

\subsection{Calculations}

In order to assist in evaluating the reliability of our results,
in this Section we make a similar analysis of the predictions
for the Sivers asymmetries obtained using a Gaussian form factor
at the proton-quark-diquark vertex given by
\begin{equation}
g(k^2)= {\sqrt{N_{R}}}\ \, (k^2-m^2)\, e^{{1\over 2 \Lambda_1^2}(k^2-\Lambda^2)} \ ,
\label{g1}
\end{equation}
where $-k^2={{\bf k}_{\perp}^2\over 1-x}+{x\over 1-x}M_{R}^2-xM^2$.
In this case, we have the distribution functions
\begin{equation}
f_1^R(x,{\bf k}_{\perp})=
{N_R\over \Lambda_1^2(1-x)}\ f_0(x)\
[(xM+m)^2+{\bf k}_{\perp}^2]\
e^{-{{\bf k}_{\perp}^2\over \Lambda_1^2(1-x)}}\ ,
\label{sa4g}
\end{equation}
where
\begin{equation}
f_0(x)=e^{x(1-x)M^2-xM_R^2\over \Lambda_1^2(1-x)}.
\label{f0}
\end{equation}
After integration over ${\bf k}_{\perp}$, we obtain the distribution functions
\begin{equation}
f_1^R(x)= \pi N_R\ f_0(x)\
[(xM+m)^2+\Lambda_1^2(1-x)]\ ,
\label{sa4gint}
\end{equation}
where the normalization factors $N_R$ are fixed by
\begin{equation}
\int_0^1 dx f_1^R(x)= 1\ .
\label{normsagau}
\end{equation}
We obtain from (\ref{app6gen}) for the present Gaussian form factor
\begin{equation}
R(x,{\bf k}_{\perp})\times
e^{-{{\bf k}_{\perp}^2\over 2 \Lambda_1^2(1-x)}}
={1\over {\bf k}_{\perp}^2}\,
\int_0^{{\bf k}_{\perp}^2}d({\bf l}_{\perp}^2)\,
e^{-{{\bf l}_{\perp}^2\over 2 \Lambda_1^2(1-x)}}
={1\over {\bf k}_{\perp}^2}\,
2 \Lambda_1^2(1-x)\,
\Big( 1- e^{-{{\bf k}_{\perp}^2\over 2 \Lambda_1^2(1-x)}}\Big)\ .
\label{R1gaussian}
\end{equation}
Finally, we obtain the following Sivers distribution functions
for the Gaussian form factor:
\begin{equation}
f_{1T}^{\perp R}(x,{\bf k}_{\perp})=
a_R\
{e_1e_2\over 4\pi}\ N_R\
2M(xM+m)\ f_0(x)\
{1\over {\bf k}_{\perp}^2}\ e^{-{{\bf k}_{\perp}^2\over 2 \Lambda_1^2(1-x)}}\
\Big(\ 1\ -\ e^{-{{\bf k}_{\perp}^2\over 2 \Lambda_1^2(1-x)}}\ \Big)\ ,
\qquad
\label{sa6g}
\end{equation}
where $a_s=1$ and $a_a=-{1\over 3}$ as in (\ref{sa6}).
Relevant formulae and details of the derivation and the relation between the light-cone and
rest-frame SU(6) descriptions of the nucleon wavefunction are given in the Appendix.

As one can see in the top panel of Fig.~\ref{Fig:kmax_gauss},
the convergence of the
${\bf k_\perp}$ integration is faster with the Gaussian form factor
than with the dipole form factor, and
in this section we use $|{\bf k}_{\perp}|_{\rm max}=1.5$ GeV
for the upper limit of the ${\bf k}_{\perp}$ integration. In the
middle panel of Fig.~\ref{Fig:kmax_gauss}, we delineate the
regions of the $(x, {\bf k}_{\perp})$ where the naively calculated
one-gluon-exchange FSI contribution to the Sivers asymmetry
in scattering off a $u$ quark exceeds unity (white), and also indicate
the magnitudes calculated at lower values of $x$ and ${\bf k}_{\perp}$.
The bottom panel presents the same information for a $d$ quark.
As in the case of the dipole form factor the unitarity cut of the ${\bf k}_{\perp}$
integration limit alters our predictions by amounts
that are again far smaller than the experimental errors.

The bottom panel presents the same information for a $d$ quark.
As in the case of the dipole form factor, we cut off the ${\bf k}_{\perp}$
integral at the solid line, which alters our predictions by amounts
that are again far smaller than the experimental errors.

\begin{figure}
\begin{center}
\includegraphics[width=1.\columnwidth]{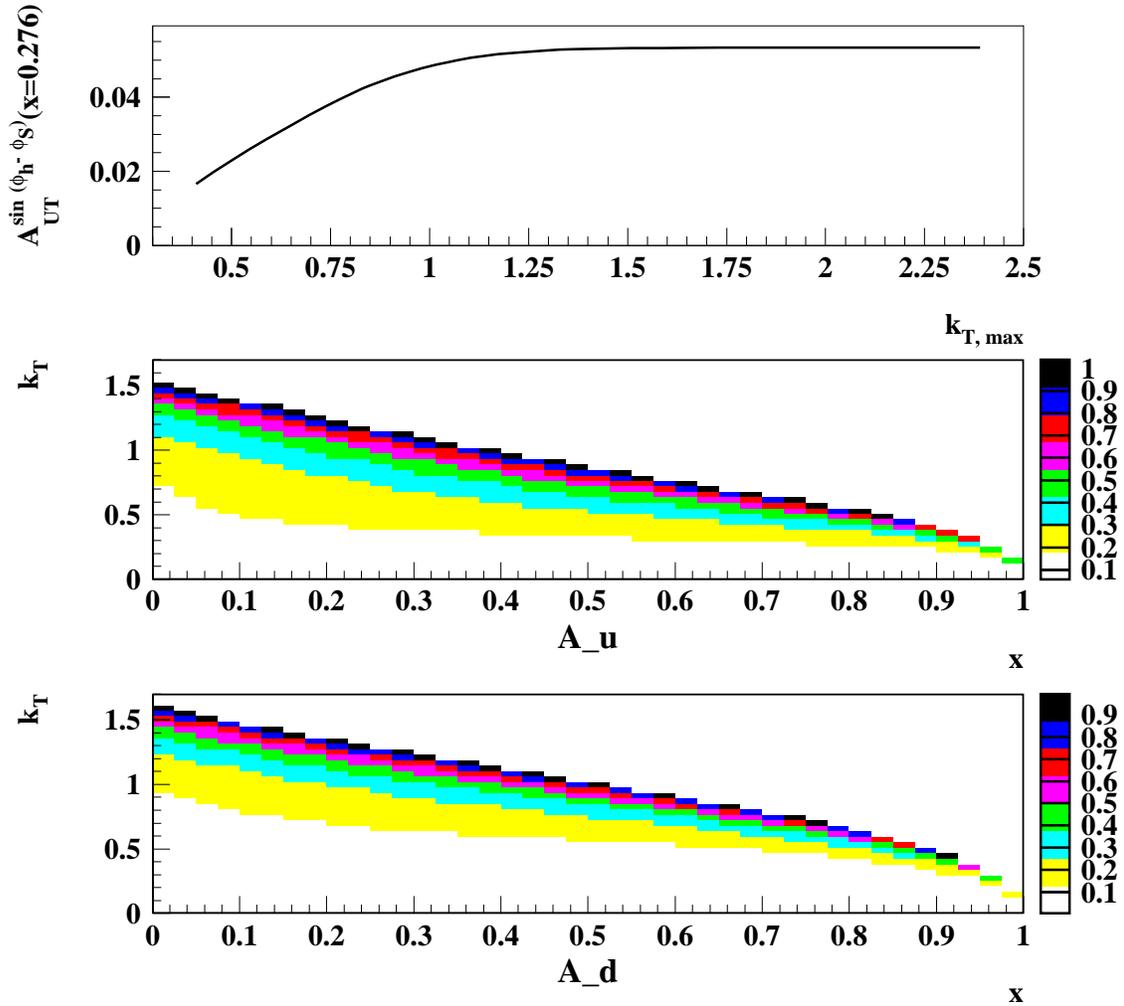}
\end{center}
\caption{\it In the top panel,
we show the Sivers asymmetry for $\pi^+$ production off a proton target
for HERMES kinematics at $x=0.276$ as a function of the upper
limit of the $|{\bf k}_{\perp}|$ integration.
The analyzing power of Sivers function for scattering off a $u$ quark (middle panel) and
a $d$ quark (bottom panel)as functions of $x$ and $k_T=|{\bf k}_{\perp}|$. The values exceed the unitarity limit in the white regions at larger $x$ and $k_T$.
All these plots are obtained using Gaussian form factors
with parameters $\Lambda_1=0.5$ GeV and $\alpha_s=0.3$.}
\label{Fig:kmax_gauss}
\end{figure}

\begin{figure}
\includegraphics[width=1.\columnwidth]{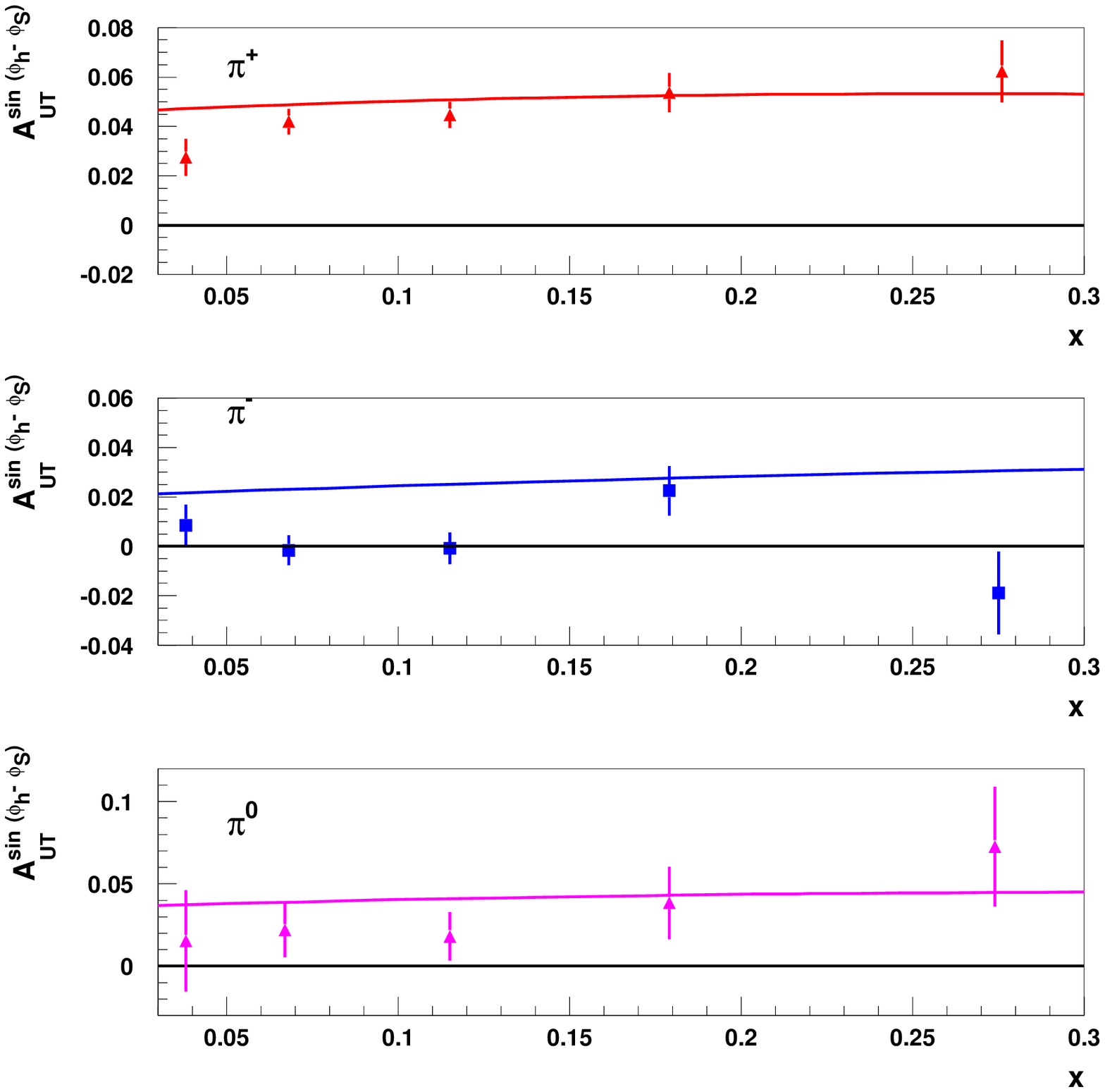}
\caption{\it Comparisons of our predictions for the Sivers asymmetries in the
production of $\pi^+$ (top panel), $\pi^-$ (middle panel) and $\pi^0$
(bottom panel) with HERMES data, assuming $\alpha_s=0.3$ and $\Lambda_1=0.5$ GeV
for the Gaussian model parameters.}
\label{Fig:h_gauss03045}
\end{figure}

\begin{figure}
\vspace{1.0cm}
\begin{center}
\includegraphics[width=1.05\columnwidth]{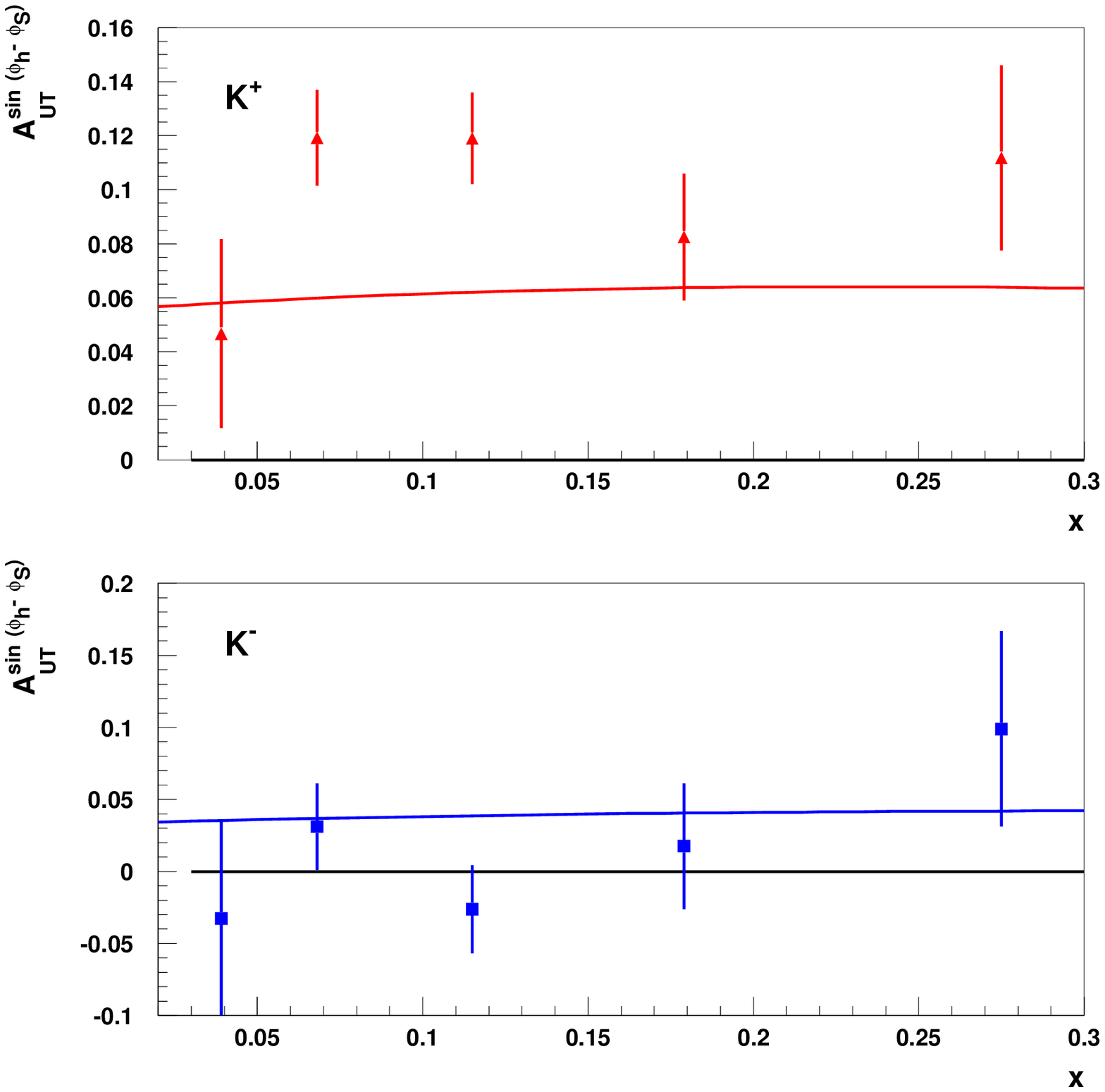}\\
\caption{\it Comparisons of our predictions for the Sivers asymmetries in the
production of $K^+$ (top panel) and $K^-$
(bottom panel) with HERMES data, assuming $\alpha_s=0.3$ and $\Lambda_1=0.5$ GeV
for the Gaussian model parameters.}
\label{Fig:kh_gauss03045}
\end{center}
\vspace{1.5cm}
\end{figure}

\begin{figure}
\includegraphics[width=1.\columnwidth]{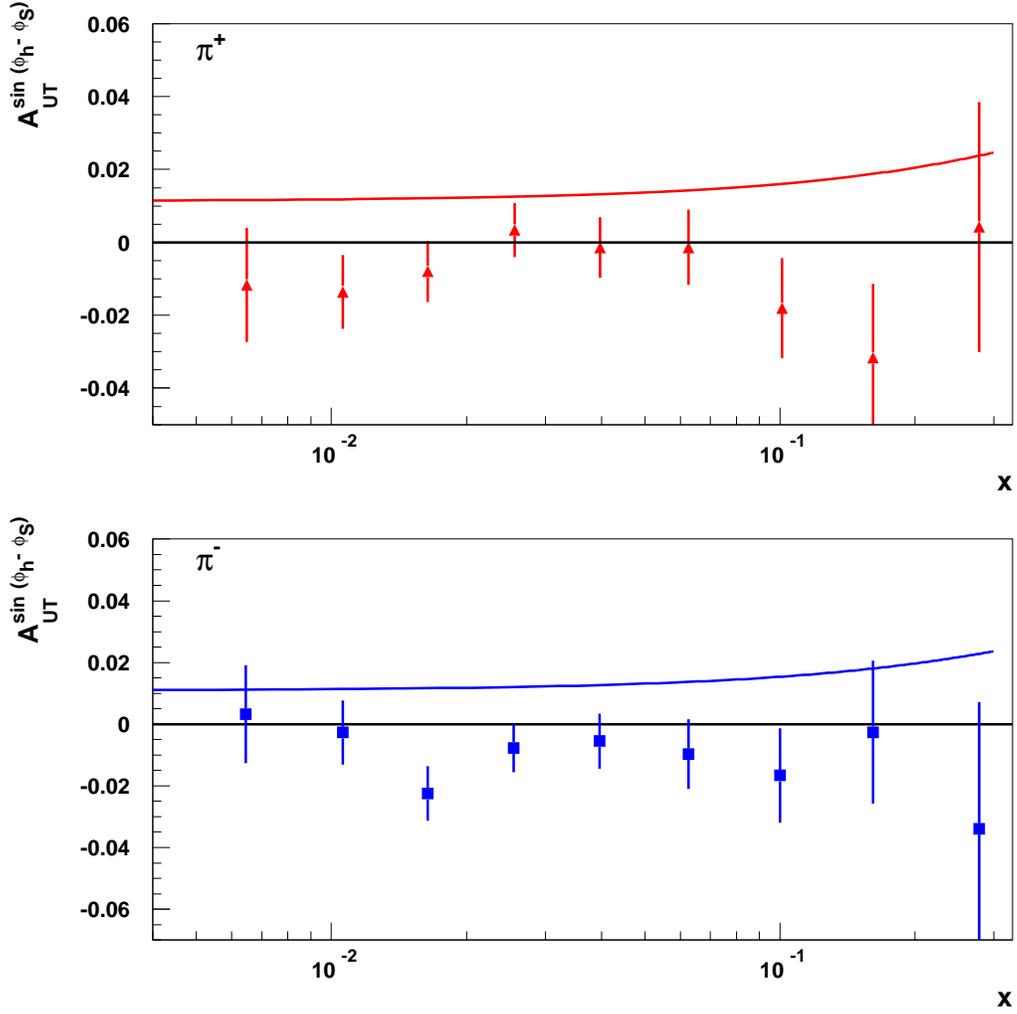}
\caption{\it Comparisons of our predictions for the Sivers asymmetries in the
production of $\pi^+$ (top panel), $\pi^-$ (bottom panel) with COMPASS deuteron data for $\pi^\pm$, assuming $\alpha_s=0.3$ and $\Lambda_1=0.5$ GeV for the Gaussian model parameters.}
\label{Fig:c_gauss03045}
\end{figure}

\begin{figure}
\vspace{1.5cm}
\begin{center}
\includegraphics[width=1.08\columnwidth]{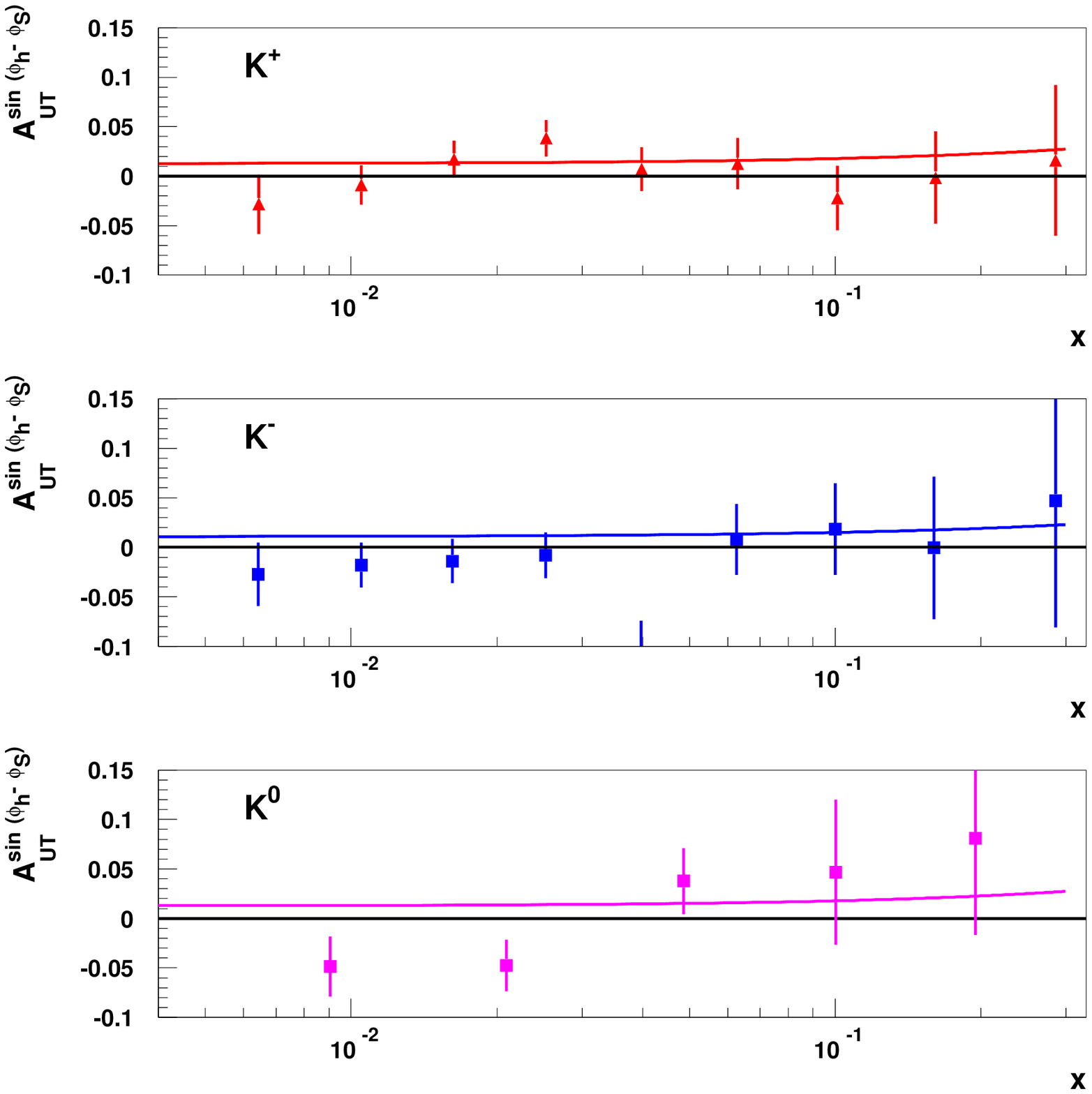}\\
\caption{\it Comparisons of our predictions for the Sivers asymmetries in the
production of $K^+$ (top panel), $K^-$ (middle panel) and $K^0$
(bottom panel) with COMPASS deuteron data, assuming $\alpha_s=0.3$ and $\Lambda_1=0.5$ GeV
for the Gaussian model parameters.}
\label{Fig:kc_gauss03045}
\end{center}
\vspace{1.0cm}
\end{figure}

\subsection{Comparisons with Data}

We display in Figs. \ref{Fig:h_gauss03045} and \ref{Fig:kh_gauss03045}
comparisons of our predictions for the Sivers asymmetries with HERMES data on pion and
kaon production, respectively. These predictions are calculated with
the Gaussian form factor using $\alpha_s=0.3$, $\Lambda_1=0.5$ GeV.
As in the dipole case, we find quantitative success for the $\pi^+$
asymmetry and qualitative success for the $\pi^0$ asymmetry, while the
experimental errors in the $\pi^-$ asymmetry do not permit a firm conclusion to be drawn.
In the case of the $K^\pm$ asymmetries, we again find a qualitative success for
the $K^+$ case, though the measured asymmetry is larger than our prediction.
This may again be explained by ignorance of the ${\bar s}$ contribution.
In the $K^-$ case, the prediction is also qualitatively successful, though no
definite conclusion can be drawn.

The corresponding comparisons between our predictions and COMPASS data
on pion and kaon production are shown in Figs.~\ref{Fig:c_gauss03045} and
\ref{Fig:kc_gauss03045}, respectively. In the $\pi^\pm$ cases we predict smaller
Sivers asymmetries than for HERMES, and the data certainly reflect this trend,
though the data are quite compatible with zero.
In the cases of the kaon asymmetries shown in
Fig.~\ref{Fig:kc_gauss03045}, the predicted asymmetries are very close to zero,
as are the values measured by COMPASS.

We see that our model gives qualitatively successful predictions for
the Sivers asymmetries also if a Gaussian form factor is assumed, and that
the predictions using this and a dipole form factor are qualitatively similar.
This gives some further confidence in the stability of our results and the conclusions we draw.

\section{Conclusions}

We have studied the Sivers single-spin asymmetry in SIDIS generated
by the mechanism of a one-gluon-exchange final-state interaction.
We derived a general formula that can be used to calculate the Sivers distribution
function for diquark models having different form factors at the
nucleon-quark-diquark vertex. We calculated the Sivers distribution functions in
diquark models with both dipole and Gaussian form factors, and compared the
corresponding predictions for Sivers single-spin asymmetries in pion and kaon
production in SIDIS with the results of HERMES and COMPASS. The predictions
made using dipole and Gaussian form factors are quite similar, and are
relatively insensitive to the unphysical values of the model calculations at
large $x$ and ${\bf k}_{\perp}$.

We find qualitatively successful results for the asymmetries in
$\pi^+, \pi^0$ and $K^+$ production. In the case of $K^+$ production at HERMES,
the measured values are even larger than our predictions, reflecting the
possible importance of an ${\bar s}$ contribution. In other cases, particularly at
COMPASS energies, many of the experimental measurements are currently
compatible with zero, and greater accuracy will be necessary to confront our
theoretical predictions. On the theoretical side, it is desirable to improve the
accuracy of our predictions, in particular by going beyond the simple
one-gluon-exchange final-state interaction. However, the relative success of
this first confrontation between HERMES and COMPASS data and our naive
predictions is an encouraging indication that one may be able to understand
satisfactorily the Sivers asymmetries in SIDIS, which are rather subtle aspects
of hadron dynamics in deep-inelastic scattering.

One {\it prima facie} conclusion from the successful comparison between our
predictions and the HERMES data on $\pi^+, \pi^0$ and $K^+$ production is
that the quark partons in the nucleon must have nonzero orbital angular momentum
in the infinite-momentum frame.

\section*{Acknowledgements}
We thank Alessandro Bacchetta, Stan Brodsky, Leonard Gamberg, Piet Mulders,
Marco Radici and Misha Sapozhnikov for helpful discussions. 
This work was supported in part by the International Cooperation
Program of the KICOS (Korea Foundation for International Cooperation
of Science \& Technology).

\vspace{1.0cm}

\section*{Appendix: Wavefunctions of Scalar and
Axial-vector Diquark Models}

The expansion of the proton state in terms of light-cone Fock states
is
\begin{eqnarray}
\left\vert \psi_p(P^+, {\bf P}_\perp )\right> &=& \sum_{n}\
\prod_{i=1}^{n}
  {{d}x_i\, {d}^2 {\bf k}_{\perp i}
\over \sqrt{x_i}\, 16\pi^3}\ \,
16\pi^3 \delta\left(1-\sum_{i=1}^{n} x_i\right)\,
\delta^{(2)}\left(\sum_{i=1}^{n} {\bf k}_{\perp i}\right)
\label{a318}
\\
&& \qquad \rule{0pt}{4.5ex}
\times \psi_n(x_i,{\bf k}_{\perp i},
\lambda_i) \left\vert n;\,
x_i P^+, x_i {\bf P}_\perp + {\bf k}_{\perp i}, \lambda_i\right>,
\nonumber
\end{eqnarray}
where the light-cone momentum fractions $x_i = k^+_i/P^+$ and ${\bf
k}_{\perp i}$ represent the momenta of the QCD
constituents. The physical transverse momenta are ${\bf p}_{\perp i}
= x_i {\bf P}_\perp + {\bf k}_{\perp i}.$ The $\lambda_i$ label the
light-cone spin projections $S^z$ of the quarks and gluons along the
quantization direction $z$. The $n$-particle states are normalized as
\begin{equation}
\left< n;\, p'_i{}^+, {\bf p}\,'_{\perp i}, \lambda'_i \right. \,
\left\vert n;\,
p^{~}_i{}^{\!\!+}, {\bf p}^{~}_{\perp i}, \lambda_i\right>
= \prod_{i=1}^n 16\pi^3
  p_i^+ \delta(p'_i{}^{+} - p^{~}_i{}^{\!\!+})\
  \delta^{(2)}( {\bf p}\,'_{\perp i} - {\bf p}^{~}_{\perp i})\
  \delta_{\lambda'_i \lambda^{~}_i}\ .
\label{normalize}
\end{equation}
Here and in the following we do not display the other quantum
numbers of the partons, {i.e.}, color and quark flavor.

\begin{figure}[h!]
\begin{center}
\vspace{0.5cm}
\includegraphics[width=0.75\columnwidth]{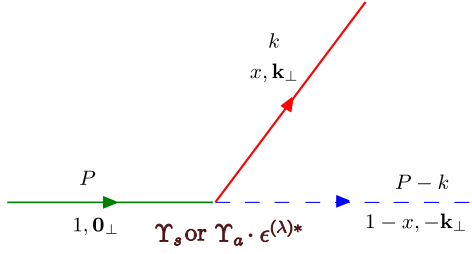}\\
\vspace{-1.0cm}
\caption{\it Diagram giving the light-cone wavefunctions of scalar and
axial-vector diquark models.}
\label{fig:waveftnsa}
\end{center}
\end{figure}

In order to construct diquark models, we take
the following form factors
at the proton-quark-diquark vertex, for the scalar ($s$) and axial-vector ($a$)
diquarks respectively:
\begin{equation}
\Upsilon_s=g_s(k^2)\ , \qquad
\Upsilon_a^\mu ={g_a(k^2)\over {\sqrt{3}}}\gamma^\mu\gamma_5 \ .
\label{sa1}
\end{equation}
We can then obtain the light-cone wavefunctions of scalar and
axial-vector diquark models from Fig. \ref{fig:waveftnsa}.
In this Appendix we consider elementary vertices given by
$g_s(k^2)=g_a(k^2)=1$.
In Sections \ref{sec:dipole} and \ref{sec:gaussian} we introduce
dipole and Gaussian form factors, respectively.

\subsection*{A. Scalar Diquark Model}

We also use the term `Yukawa model' for the scalar diquark model
described in this Subsection.
Each Fock-state wavefunction of the physical proton with total spin
projection $J^z = \pm {1\over 2}$ is represented by a function
$\psi^{J^z}_n(x_i,{\bf k}_{\perp i},\lambda_i)$, where
\begin{equation}
k_i=(k^+_i,{\bf k}_{ \perp i},k^-_i)=
\left(x_i P^+, {\bf k}^{~}_{\perp i},
\frac{{\bf k}_{\perp i}^2+m_i^2}{x_i P^+}\right)
\end{equation}
specifies the momentum of each constituent and $\lambda_i$ specifies
its light-cone helicity in the $z$ direction.

From Fig. \ref{fig:waveftnsa} with the scalar vertex $\Upsilon_s$,
the $J^z = + {1\over 2}$ two-particle Fock state
in the scalar diquark model is given
by~\cite{BD80,BHMS01}
\begin{eqnarray}
&&\left|\Psi^{\uparrow}_{\rm two \ particle}(P^+, {\bf P}_\perp = {\bf
0}_\perp)\right>
\label{sn1}\\
&=&
\int\frac{{d} x \, {d}^2
           {\bf k}_{\perp} }{\sqrt{x(1-x)}\, 16 \pi^3}
\Big[ \
\psi^{\uparrow}_{+\frac{1}{2}} (x,{\bf k}_{\perp})\,
\left| +\frac{1}{2}\, ;\,\, xP^+\, ,\,\, {\bf k}_{\perp} \right>
+\psi^{\uparrow}_{-\frac{1}{2}} (x,{\bf k}_{\perp})\,
\left| -\frac{1}{2}\, ;\,\, xP^+\, ,\,\, {\bf k}_{\perp} \right>\ \Big]\ ,
\nonumber
\end{eqnarray}
where
\begin{equation}
\left
\{ \begin{array}{l}
\psi^{\uparrow}_{+\frac{1}{2}} (x,{\bf k}_{\perp})=\frac{(xM+m)}{x }\,
\varphi \ ,\\
\psi^{\uparrow}_{-\frac{1}{2}} (x,{\bf k}_{\perp})=
-\frac{(+k^1+{\mathrm i} k^2)}{x }\, \varphi
\ .
\end{array}
\right.
\label{sn2}
\end{equation}
The scalar part of the wavefunction $\varphi$ is given by \cite{BD80,BHMS01}
\begin{equation}
\varphi (x,{\bf k}_{\perp}) = \frac{g}{\sqrt{1-x}}\
\frac{1}{M^2-{{\bf k}_{\perp}^2+m^2 \over x}
-{{\bf k}_{\perp}^2+\lambda^2 \over 1-x}}
=-g{x{\sqrt{1-x}}\over {\bf k}_{\perp}^2+B}
\ ,
\label{wfdenom}
\end{equation}
where
\begin{equation}
B=x(1-x)\Bigl( -M^2+{m^2\over x}+{{\lambda}^2\over 1-x} \Bigr)\ .
\label{wfdenom2}
\end{equation}
Similarly, the $J^z = - {1\over 2}$ two-particle Fock state is given by
\cite{BD80,BHMS01}
\begin{eqnarray}
&&\left|\Psi^{\downarrow}_{\rm two \ particle}(P^+, \bf P_\perp =
\bf 0_\perp)\right>
\label{sn1a}\\
&=&
\int\frac{{d} x \, {d}^2
           {\bf k}_{\perp} }{\sqrt{x(1-x)}\, 16 \pi^3}
\Big[ \
\psi^{\downarrow}_{+\frac{1}{2}} (x,{\bf k}_{\perp})\,
\left| +\frac{1}{2}\, ;\,\, xP^+\, ,\,\, {\bf k}_{\perp} \right>
+\psi^{\downarrow}_{-\frac{1}{2}} (x,{\bf k}_{\perp})\,
\left| -\frac{1}{2}\, ;\,\, xP^+\, ,\,\, {\bf k}_{\perp} \right>\ \Big]\ ,
\nonumber
\end{eqnarray}
where
\begin{equation}
\left
\{ \begin{array}{l}
\psi^{\downarrow}_{+\frac{1}{2}} (x,{\bf k}_{\perp})=
-\frac{(-k^1+{\mathrm i} k^2)}{x }\,
\varphi \ ,\\
\psi^{\downarrow}_{-\frac{1}{2}} (x,{\bf k}_{\perp})=\frac{(xM+m)}{x}\,
\varphi
\ .
\end{array}
\right.
\label{sn2a}
\end{equation}

The quark-state bases on the right-hand sides of (\ref{sn1}) and (\ref{sn1a})
correspond to light-cone spinors.
The Bjorken-Drell spinors $u^{\rm BD}(k)$ and
the light-cone spinors $u^{\rm LC}(k)$ are related by
\begin{eqnarray}
u_{+{1\over 2}}^{\rm BD}(k)&=&{1\over {\sqrt{(k^++m)^2+{\vec k}_{\perp}^2}}}
\ \Bigl(\ (k^++m)\ u_{+{1\over 2}}^{\rm LC}(k)\ -\
(k^1+ik^2)\ u_{-{1\over 2}}^{\rm LC}(k)\ \Bigr)\ ,
\nonumber\\
u_{-{1\over 2}}^{\rm BD}(k)&=&{1\over {\sqrt{(k^++m)^2+{\vec k}_{\perp}^2}}}
\ \Bigl(\
-(-k^1+ik^2)\ u_{+{1\over 2}}^{\rm LC}(k)\ +\
(k^++m)\ u_{-{1\over 2}}^{\rm LC}(k)\ \Bigr)\ .
\qquad
\label{sas4}
\end{eqnarray}
The Bjorken-Drell spinors
$u^{\rm BD}(k)$ satisfy
\begin{equation}
j^+u_{+{1\over 2}}^{\rm BD}(k)=0\ ,\ \ \
j^-u_{-{1\over 2}}^{\rm BD}(k)=0\ ,\ \ \
j^-u_{+{1\over 2}}^{\rm BD}(k)=u^{\rm BD}_{-{1\over 2}}(k)\ ,\ \ \
j^+u_{-{1\over 2}}^{\rm BD}(k)=u_{+{1\over 2}}^{\rm BD}(k)\ ,
\label{sap4}
\end{equation}
where
$j^{\pm}=j^1\pm i j^2$ and
\begin{equation}
j^i = s^i + l^i\ ,\ \ \
s^i=\Sigma^i =
\left(
\begin{array}{cc}
\sigma^i&0\\
0&\sigma^i
\end{array}
\right)
\ ,\ \ \
l^i = -i \epsilon^{ijk} k^j{\partial\over \partial k^k}\ ,\ \ \
\epsilon^{123}=1\ .
\label{sap5}
\end{equation}
Since $k^+=xM$ in the proton rest frame, from (\ref{sas4}) the Fock states given in
(\ref{sn1}) and (\ref{sn1a}) with the light-cone wavefunctions
(\ref{sn2}) and (\ref{sn2a}) correspond to
$\left| +\frac{1}{2}\, ;\,\, xP^+\, ,\,\, {\bf k}_{\perp} \right>^{\rm BD}$
and $\left| -\frac{1}{2}\, ;\,\, xP^+\, ,\,\, {\bf k}_{\perp} \right>^{\rm BD}$,
respectively, in the proton rest frame.

\subsection*{B. Axial-vector Diquark Model}

For the polarization vectors of the axial-vector diquark appearing at the
vertex of Fig. \ref{fig:waveftnsa},
we use the following set of three polarization vectors:
\begin{eqnarray}
{\epsilon}^{(+1){\mu}}&=&
({\epsilon}^{(+1)0}, {\epsilon}^{(+1)1}, {\epsilon}^{(+1)2}, {\epsilon}^{(+1)3})
\ =\ {1 \over {\sqrt{2}}} (0,-1,-i,0)\ ,
\nonumber\\
{\epsilon}^{(-1){\mu}}&=&{1 \over {\sqrt{2}}} (0,+1,-i,0)\ ,
\label{axipol}\\
{\epsilon}^{(0){\mu}}&=&({P^3 \over M},0,0,{P^0 \over M})\ .
\nonumber
\end{eqnarray}
These polarization vectors are those used in Ref. \cite{JMR97} which satisfy
$\sum_{\lambda} {\epsilon}_{\mu}^{(\lambda)*} {\epsilon}_{\nu}^{(\lambda)}
= - g_{\mu\nu} + P_{\mu}P_{\nu}/M^2$.
In the proton rest frame, ${\epsilon}^{(0){\mu}}$ is the spin vector
oriented in the $z$-direction and ${\epsilon}^{(\pm 1){\mu}}$
are those circularly polarized in the $x$-$y$ plane.

From Fig. \ref{fig:waveftnsa} with the scalar vertex
$\Upsilon_a\cdot \epsilon^{(\lambda) *}$,
the two-particle Fock state for the proton with $J^z = + {1\over 2}$ has
six possible spin combinations for the quark and axial-vector diquark:
\begin{eqnarray}
\lefteqn{
\left|\Psi^{\uparrow}_{\rm two \ particle}(P^+, \bf P_\perp = \bf
0_\perp)\right> =
\int\frac{{d} x \, {d}^2
           {\bf k}_{\perp} }{\sqrt{x(1-x)}\, 16 \pi^3}
}
\label{vsn1}\\
&&
\left[ \ \ \,
\psi^{\uparrow}_{+\frac{1}{2}\, +1}(x,{\bf k}_{\perp})\,
\left| +\frac{1}{2}\, +1\, ;\,\, xP^+\, ,\,\, {\bf k}_{\perp}\right>
+\psi^{\uparrow}_{-\frac{1}{2}\, +1}(x,{\bf k}_{\perp})\,
\left| -\frac{1}{2}\, +1\, ;\,\, xP^+\, ,\,\, {\bf k}_{\perp}\right>
\right.
\nonumber\\
&&\left. {}
+\psi^{\uparrow}_{+\frac{1}{2}\, 0} (x,{\bf k}_{\perp})\,
\left| +\frac{1}{2}\, \ \ 0\, ;\,\, xP^+\, ,\,\, {\bf k}_{\perp}\right>
+\psi^{\uparrow}_{-\frac{1}{2}\, 0} (x,{\bf k}_{\perp})\,
\left| -\frac{1}{2}\, \ \ 0\, ;\,\, xP^+\, ,\,\, {\bf k}_{\perp}\right>
\right.
\nonumber\\
&&\left. {}
+\psi^{\uparrow}_{+\frac{1}{2}\, -1} (x,{\bf k}_{\perp})\,
\left| +\frac{1}{2}\, -1\, ;\,\, xP^+\, ,\,\, {\bf k}_{\perp}\right>
+\psi^{\uparrow}_{-\frac{1}{2}\, -1} (x,{\bf k}_{\perp})\,
\left| -\frac{1}{2}\, -1\, ;\,\, xP^+\, ,\,\, {\bf k}_{\perp}\right>\
\right] \ ,
\nonumber
\end{eqnarray}
where the two-particle states $|s_{\rm f}^z, s_{\rm b}^z; \ x, {\bf
k}_{\perp} \rangle$ are normalized as in (\ref{normalize}). Here $s_{\rm
f}^z$ and $s_{\rm b}^z$ denote the $z$-component of the spins of the
constituent fermion and boson, respectively, and the variables $x$ and
${\bf k}_{\perp}$ refer to the momentum of the fermion. The
wavefunctions are given by
\begin{equation}
\left
\{ \begin{array}{l}
\psi^{\uparrow}_{+\frac{1}{2}\, +1} (x,{\bf k}_{\perp})=-{\sqrt{{2\over 3}}}
\ \frac{(-k^1+{i} k^2)}{x}\,
\varphi \ ,\\
\psi^{\uparrow}_{-\frac{1}{2}\, +1} (x,{\bf k}_{\perp})=+{\sqrt{{2\over 3}}}
\ {(xM+m)\over x}\,
\varphi \ ,\\
\psi^{\uparrow}_{+\frac{1}{2}\, \ 0} (x,{\bf k}_{\perp})=-{\sqrt{{1\over 3}}}
\ {(xM+m)\over x}\,
\varphi \ ,\\
\psi^{\uparrow}_{-\frac{1}{2}\, \ 0} (x,{\bf k}_{\perp})=+{\sqrt{{1\over 3}}}
\ \frac{(+k^1+{i} k^2)}{x}\,
\varphi \ ,\\
\psi^{\uparrow}_{+\frac{1}{2}\, -1} (x,{\bf k}_{\perp})=0\ ,\\
\psi^{\uparrow}_{-\frac{1}{2}\, -1} (x,{\bf k}_{\perp})=0\ ,
\end{array}
\right.
\label{vsn2}
\end{equation}
where
\begin{equation}
\varphi (x,{\bf k}_{\perp}) = \frac{e}{\sqrt{1-x}}\
\frac{1}{M^2-{{\bf k}_{\perp}^2+m^2 \over x}
-{{\bf k}_{\perp}^2+\lambda^2 \over 1-x}}\ .
\label{wfdenomaxial}
\end{equation}
Similarly, the wavefunctions for a proton with negative helicity
are given by
\begin{equation}
\left
\{ \begin{array}{l}
\psi^{\downarrow}_{+\frac{1}{2}\, +1} (x,{\bf k}_{\perp})=0\ ,\\
\psi^{\downarrow}_{-\frac{1}{2}\, +1} (x,{\bf k}_{\perp})=0\ ,\\
\psi^{\downarrow}_{+\frac{1}{2}\, \ 0} (x,{\bf k}_{\perp})=-{\sqrt{{1\over 3}}}
\ \frac{(-k^1+{i} k^2)}{x}\,
\varphi \ ,\\
\psi^{\downarrow}_{-\frac{1}{2}\, \ 0} (x,{\bf k}_{\perp})=+{\sqrt{{1\over 3}}}
\ {(xM+m)\over x}\,
\varphi \ ,\\
\psi^{\downarrow}_{+\frac{1}{2}\, -1} (x,{\bf k}_{\perp})=-{\sqrt{{2\over 3}}}
\ {(xM+m)\over x}\,
\varphi \ ,\\
\psi^{\downarrow}_{-\frac{1}{2}\, -1} (x,{\bf k}_{\perp})=+{\sqrt{{2\over 3}}}
\ \frac{(+k^1+{i} k^2)}{x}\,
\varphi \ .
\end{array}
\right.
\label{vsn2a}
\end{equation}
The coefficients of $\varphi$ in (\ref{vsn2})
and (\ref{vsn2a}) are the matrix elements of
$\frac{\overline{u}(k^+,k^-,{\bf k}_{\perp})}{{\sqrt{k^+}}}
\, \Upsilon_a \cdot \epsilon^{(\lambda) *}\,
\frac{u (P^+,P^-,{\bf P}_{\perp})}{{\sqrt{P^+}}}$
which are
the numerators of the wavefunctions corresponding to
each constituent-spin $s^z$ configuration.

Because of the transformation properties (\ref{sap4}) of
the Bjorken-Drell spinors $u^{\rm BD}(k)$,
when we construct the proton spin states $|P; \lambda =\pm {1\over 2}\rangle$
by using the Clebsch-Gordan coefficients with $u^{\rm BD}(k)$ for
spin-half components, they satisfy
\begin{eqnarray}
&&{J}^+|P; \lambda =+{1\over 2}\rangle=0\ ,\ \ \
{J}^-|P; \lambda =-{1\over 2}\rangle=0
\nonumber\\
&&{J}^-|P; \lambda =+{1\over 2}\rangle=|P; \lambda =-{1\over 2}\rangle\ ,
\ \ \
{J}^+|P; \lambda =-{1\over 2}\rangle=|P; \lambda =+{1\over 2}\rangle\ ,
\label{p4a}
\end{eqnarray}
where ${J}^i$ is the angular momentum operator for the proton given by
${J}^i=\sum_a j^i_a =\sum_a ({s}^i_a+{l}^i_a)$,
which is the sum over the constituents $a$,
and $J^{\pm}=J^1\pm i J^2$.

Since $k^+=xM$ in the proton rest frame,
from (\ref{sas4})
the Fock states with the light-cone wavefunctions given in (\ref{vsn2})
and (\ref{vsn2a}) correspond to the following states in the proton rest frame,
respectively:
$$\Big( {\sqrt{2\over 3}}
\left| -\frac{1}{2}\, +1\, ;\,\, xP^+\, ,\,\, {\bf k}_{\perp}\right>^{\rm BD}
-{\sqrt{1\over 3}}
\left| +\frac{1}{2}\, \ 0\, ;\,\, xP^+\, ,\,\, {\bf k}_{\perp}\right>^{\rm BD}
\Big)\ ,$$
$$\Big( -{\sqrt{2\over 3}}
\left| +\frac{1}{2}\, -1\, ;\,\, xP^+\, ,\,\, {\bf k}_{\perp}\right>^{\rm BD}
+{\sqrt{1\over 3}}
\left| -\frac{1}{2}\, \ 0\, ;\,\, xP^+\, ,\,\, {\bf k}_{\perp}\right>^{\rm BD}
\Big)\ .$$
The above states are angular momentum eigenstates
$|P; \lambda =+{1\over 2}\rangle$ and $|P; \lambda =-{1\over 2}\rangle$, respectively,
which satisfy (\ref{p4a}).
In the proton rest frame,
the states described by (\ref{vsn2}) and ((\ref{vsn2a}) are identical to
the axial-vector diquark states appearing in the SU(6) wavefunctions
(\ref{su6p1}) and (\ref{su6p2}). Therefore, in the proton rest frame the
wavefunctions of the axial-vector diquark states derived here
through the vertex given in Fig. \ref{fig:waveftnsa} with the polarization
vectors given in (\ref{axipol}) coincide with those in the SU(6)
wavefunction for the proton.


\newpage
\begin{thebibliography}{99}

\bibitem{Sivers}
  D.W.~Sivers,
  Phys.\ Rev.\  D {\bf 41}, 83 (1990);
  Phys.\ Rev.\  D {\bf 43}, 261 (1991).

\bibitem{Collins:1992kk}
  J.C.~Collins,
  Nucl.\ Phys.\  B {\bf 396}, 161 (1993).

\bibitem{Anselmino:1994tv}
  M.~Anselmino, M.~Boglione, and F.~Murgia,
  Phys.\ Lett.\  B {\bf 362}, 164 (1995).

\bibitem{BM98} D. Boer and P.J. Mulders, Phys. Rev. D {\bf 57}, 5780 (1998).


\bibitem{Adams}
  D.L.~Adams et al. [FNAL-E704 Collaboration],
  Phys.\ Lett.\  B {\bf 264}, 462 (1991).

\bibitem{BHS} S.J. Brodsky, D.S. Hwang, and I. Schmidt,
Phys. Lett. B {\bf 530}, 99 (2002).


\bibitem{Airapetian:2004tw}
  A.~Airapetian et al. [HERMES Collaboration],
  Phys.\ Rev.\ Lett.\  {\bf 94}, 012002 (2005).

\bibitem{HERMESDIS2007} M. Diefenthaler [HERMES Collaboration],
arXiv:0706.2242 [hep-ex],
in the proceedings of 15th International Workshop on Deep-Inelastic Scattering and
Related Subjects (DIS2007), 579-582, Munich, Germany, 16-20 Apr 2007.

\bibitem{COMPASS0802} M. Alekseev et al. [COMPASS Collaboration],
arXiv:0802.2160 [hep-ex].

\bibitem{Vogelsang05} W. Vogelsang and F. Yuan, Phys. Rev. D {\bf 72}, 054028 (2005).

\bibitem{Collins06} J.C. Collins,  A.V. Efremov, K. Goeke, S. Menzel, A. Metz, and P. Schweitzer,
Phys. Rev. D {\bf 73}, 014021 (2006).

\bibitem{Arnold08} S. Arnold, A.V. Efremov, K. Goeke, M. Schlegel, and P. Schweitzer,
arXiv:0805.2137 [hep-ph]

\bibitem{Anselmino0805} M. Anselmino et al., arXiv:0805.2677 [hep-ph].

\bibitem{Boer:2003tx}
  D.~Boer and W.~Vogelsang,
  Phys.\ Rev.\  D {\bf 69}, 094025 (2004).

\bibitem{Bacchetta:2005rm}
  A.~Bacchetta, C.J.~Bomhof, P.J.~Mulders, and F.~Pijlman,
  Phys.\ Rev.\  D {\bf 72}, 034030 (2005).

\bibitem{Abelev:2007ii}
  B.I.~Abelev {\it et al.}  [STAR Collaboration],
Phys. Rev. Lett. {\bf 99}, 142003 (2007).

\bibitem{MT96} P.J. Mulders and R.D. Tangerman,
Nucl. Phys. B {\bf 461}, 197 (1996), Erratum-ibid. B {\bf 484}, 538 (1997).

\bibitem{weak} S.J. Brodksy, D.S. Hwang, and I. Schmidt,
Phys. Lett. B {\bf 553}, 223 (2003).

\bibitem{JMR97} R. Jakob, P.J. Mulders, and J. Rodrigues,
Nucl. Phys. A {\bf 626}, 937 (1997).



\bibitem{collins} J.C. Collins, Phys. Lett. B {\bf 536}, 43 (2002).

\bibitem{ji1} X. Ji and F. Yuan, Phys. Lett. B {\bf 543}, 66 (2002).

\bibitem{ji2} A. Belitsky, X. Ji, and F. Yuan,
Nucl. Phys. B {\bf 656}, 165 (2003).

\bibitem{BMP03} D. Boer, P.J. Mulders, and F. Pijlman,
Nucl. Phys. B {\bf 667}, 201 (2003).


\bibitem{BBH03} D. Boer, S.J. Brodsky, and D.S. Hwang,
Phys. Rev. D {\bf 67}, 054003 (2003).

\bibitem{BSY04} A. Bacchetta, A. Sch\"afer, and J.J. Yang,
Phys. Lett. B {\bf 578}, 109 (2004).

\bibitem{BCR0807} A. Bacchetta, F. Conti, M. Radici, arXiv:0807.0323 [hep-ph].

\bibitem{GGS07} L.P. Gamberg, G.R. Goldstein, and M. Schlegel,
Phys. Rev. D {\bf 77}, 094016 (2008).

\bibitem{Anselmino:2005nn}
  M.~Anselmino, M.~Boglione, U.~D'Alesio, A.~Kotzinian, F.~Murgia, and A.~Prokudin,
  Phys.\ Rev.\  D {\bf 71}, 074006 (2005).

\bibitem{dss} D. de Florian, R. Sassot, and M. Stratmann, Phys. Rev D {\bf 75}, 114010 (2007)

\bibitem{BD80} S.J. Brodsky and S.D. Drell, Phys. Rev. D {\bf 22}, 2236 (1980).

\bibitem{BHMS01} S.J. Brodsky, D.S. Hwang, B.-Q. Ma, and I. Schmidt,
Nucl. Phys. B {\bf 593}, 311 (2001).









\end{thebibliography}
\end{document}